\documentclass[numberedappendix]{emulateapj} 

\begin{document}

\title{The intrinsic dimensionality of spectro-polarimetric data}

\author{A. Asensio Ramos\altaffilmark{1}, H. Socas-Navarro\altaffilmark{2}, A. L\'opez Ariste\altaffilmark{3} 
\& M. J. Mart\'{\i}nez Gonz\'alez\altaffilmark{1,4,5}}
\altaffiltext{1}{Instituto de Astrof\'{\i}sica de Canarias, 38205, La Laguna, Tenerife, Spain}
\altaffiltext{2}{High Altitude Observatory, National Center for Atmospheric Research, PO Box 3000, Boulder, CO 80307-300, 
USA\footnote{The National Center for Atmospheric Research is sponsored by the National Science Foundation.}}
\altaffiltext{3}{THEMIS, CNRS-UPS 853, C/ V\'{i}a L\'actea s/n, 38200, La Laguna, Tenerife, Spain}
\altaffiltext{4}{Max-Planck Institut f\"ur Sonnensystemforschung, 37191 Katlenburg-Lindau, Germany}
\altaffiltext{5}{Present address: CNRS UMR 8112-LERMA, Observatoire de Paris, Section de Meudon, 92195 Meudon, France}
\email{aasensio@iac.es}

\begin{abstract}
The amount of information available in spectro-polarimetric data is estimated. To this end, the 
intrinsic dimensionality of the data is inferred with the aid of a recently derived estimator
based on nearest-neighbor considerations and obtained applying the principle of maximum
likelihood. We show in detail that the estimator correctly captures the intrinsic dimension
of artificial datasets with known dimension. The effect of noise in the estimated 
dimension is analyzed thoroughly and we conclude that it introduces a positive bias
that needs to be accounted for. Real simultaneous spectro-polarimetric observations 
in the visible 630~nm and the near-infrared 1.5~$\mu$m spectral
regions are also investigated in  
detail, showing that the near-infrared dataset provides more
information of the physical conditions  
in the solar atmosphere than the visible dataset. Finally, we demonstrate that the amount of
information present in an observed dataset is a monotonically increasing function of the
number of available spectral lines.
\end{abstract}

\keywords{magnetic fields --- Sun: atmosphere, magnetic fields --- line: profiles --- polarization}

\section{Introduction}
High-dimensional data present difficulties when analyzing and
understanding their statistical properties. The efficiency of typical
statistical and computational methods usually degrades very fast when
the dimensionality of the problem increases, thus making the analysis
of the observed data a cumbersome or, sometimes, unfeasible task. This
fact is often referred to as the \emph{curse of dimensionality}.  The
advent of computers has permitted to face the analysis of
increasingly complex data.  These data usually exhibit an intricate
behavior and, in order to understand the underlying physics that
produces such effects, we have been forced to develop very
complicated models. Ideally, these models have to be based on physical
grounds but there seems to be no way of knowing in advance how
complicated this model has to be to correctly reproduce the
observed behavior.

In spite of their inherent complexity, the analysis of large datasets
such as those produced by modern instrumentation, indicates
that not all measured datapoints are equally
relevant for the understanding of the underlying phenomena. In other words, it 
is clear that the reason why many simplified physical models are successful in reproducing
a large amount of observations is because the data itself is not truly high-dimensional.
Based on this premise, efforts are being made to develop methods that
are capable of reducing the dimensionality of the observed datasets while still preserving
their fundamental properties. Mathematically, the idea is that, while
the original data may have a very 
large dimensionality, they are in fact confined to a small sub-region of that high-dimensional
space. In this case, we can consider that the data ``lives'' in a subspace of low dimension 
(the so-called intrinsic dimension) that
is embedded in the high-dimensional space. This lower dimension subspace
is not usually simple to describe because it often lies in a manifold whose relation with the original
high-dimensional space has to be described by a very complex non-linear (and usually unknown) function. 
In spite of the complexity, when facing a high-dimensionality dataset, it is of great interest 
to reduce the dimension of the original data prior to any modeling
effort. In this manner we can uncover
more easily the physics underlying the observations and even detect previously unknown properties 
that can be of interest.

Among the most popular methods for dimensionality reduction we find principal component 
analysis (PCA) or Karhunen-Loeve expansion. Due to its computational simplicity, it is one of 
the most widely employed methods
\cite[e.g.,][]{rees_PCA00,arturo_casini02,socas_navarro05,casini05,ferreras06}. PCA
seeks orthogonal directions in the original high-dimensional space along which
the data correlation is the largest. From a computational point of view, the method finds the
eigenvalues and eigenvectors of the covariance matrix obtained from a given dataset.
Then, the directions on the space where the
correlation is large (large eigenvalues) 
may be approximately described with only one parameter 
(a factor multiplying the associated eigenvector) and have sometimes a
physical meaning. An 
example of this can been seen in \cite{skumanich02}, who demonstrated how the eigenvectors associated
with the largest eigenvalues of the correlation matrix obtained from spectropolarimetric observations 
of a sunspot are related to fundamental physical
parameters. 
They showed that the first
eigenvector is associated with the average spectrum, the second eigenvector gives information 
about the velocity and the third eigenvector gives information about the magnetic splitting.

One of the weakest points of PCA is its linear character, because it relies only
in the information provided by second order statistics. Therefore, it cannot
efficiently describe a dataset whose embedding in the original high-dimensional space is a nonlinear manifold.
Several methods have been developed during the last years to overcome this difficulty. Among them,
we can find Locally Linear Embedding \citep[LLE,][]{lle00}, Isomap \citep{isomap00} and
Self-Organizing Maps \citep[SOM,][]{kohonen_SOM01}. These methods are very promising and have been shown to
outperform PCA when reducing the dimension of datasets that present clear nonlinearities.
Recently, a promising non-linear extension of PCA (kernel PCA) has also been developed 
\citep{kpca98}. It is based on the extension of PCA to non-linear mappings by the application of
Mercer kernels and it effectively takes into account high-order statistics from the datasets.
Another nonlinear version of PCA can be carried out with the aid of
auto-associative neural networks (AANNs) \citep[e.g.,][for applications in the inversion of Stokes profiles]{socas_navarro_aann05}.
All the previous methods are computationally expensive. AANNs require
the training of a neural network with a bottleneck hidden layer
that contains $d$ neurons, with $d$ the expected intrinsic dimension of the dataset. This training
requires a very complex non-linear optimization that can be carried
out with standard methods, such as the backpropagation algorithm
\citep{rumelhart_backprop86,werbos_backprop94}. 
Concerning KPCA, it requires the numerical diagonalization of a very large correlation matrix of size
$N \times N$ \citep{kpca98}. For large datasets, the diagonalization
poses a heavy burden in terms of computational time and memory requirements
because the correlation matrix is not sparse. 

The above-mentioned tools have been introduced recently and probably require further study in order to understand 
all their statistical and computational properties. Unfortunately, they all suffer from a very important
limitation: none of these methods is capable of giving a reliable estimation of the intrinsic 
dimension $d$ of the datasets.
When this number is known or obtained by a different method, the previous methods are able to 
yield the projection of the original dataset in
a nonlinear subspace of dimension $d$. If $d$ is close to the correct intrinsic dimension of the 
original dataset,
they usually capture the structure of the nonlinear subspace and give good results. Although it seems
of reduced importance, a good estimation of $d$ gives the key to understanding the physics
underlying in the observations. In the framework of spectropolarimetry, 
it would be desirable to find possible direct relations between the nonlinear dimensions
captured by these methods and the physical parameters employed
for the forward modeling (magnetic field strength, filling factor, macroscopic
velocities, etc.).
If $d$ is too small, important features of the data are projected
onto the same dimension and part of the information available is
lost. If, on the contrary,
$d$ is too large, then the methods can introduce noise in the nonlinear manifold. Also important is the
fact that a good estimation of $d$ is very important to reduce the computational work and avoid a
trial-and-error procedure.

\begin{figure*}[!ht]
 \plottwo{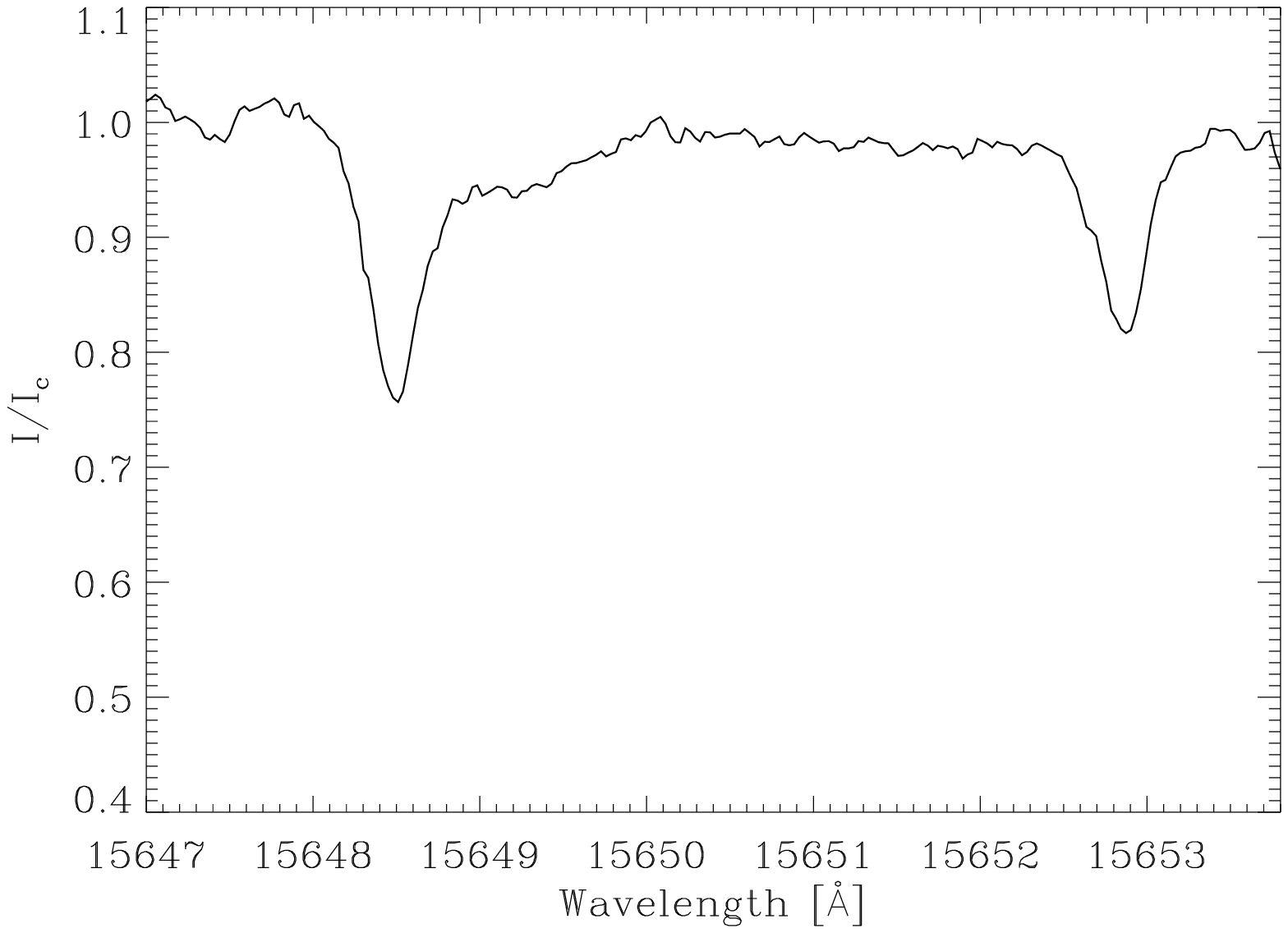}{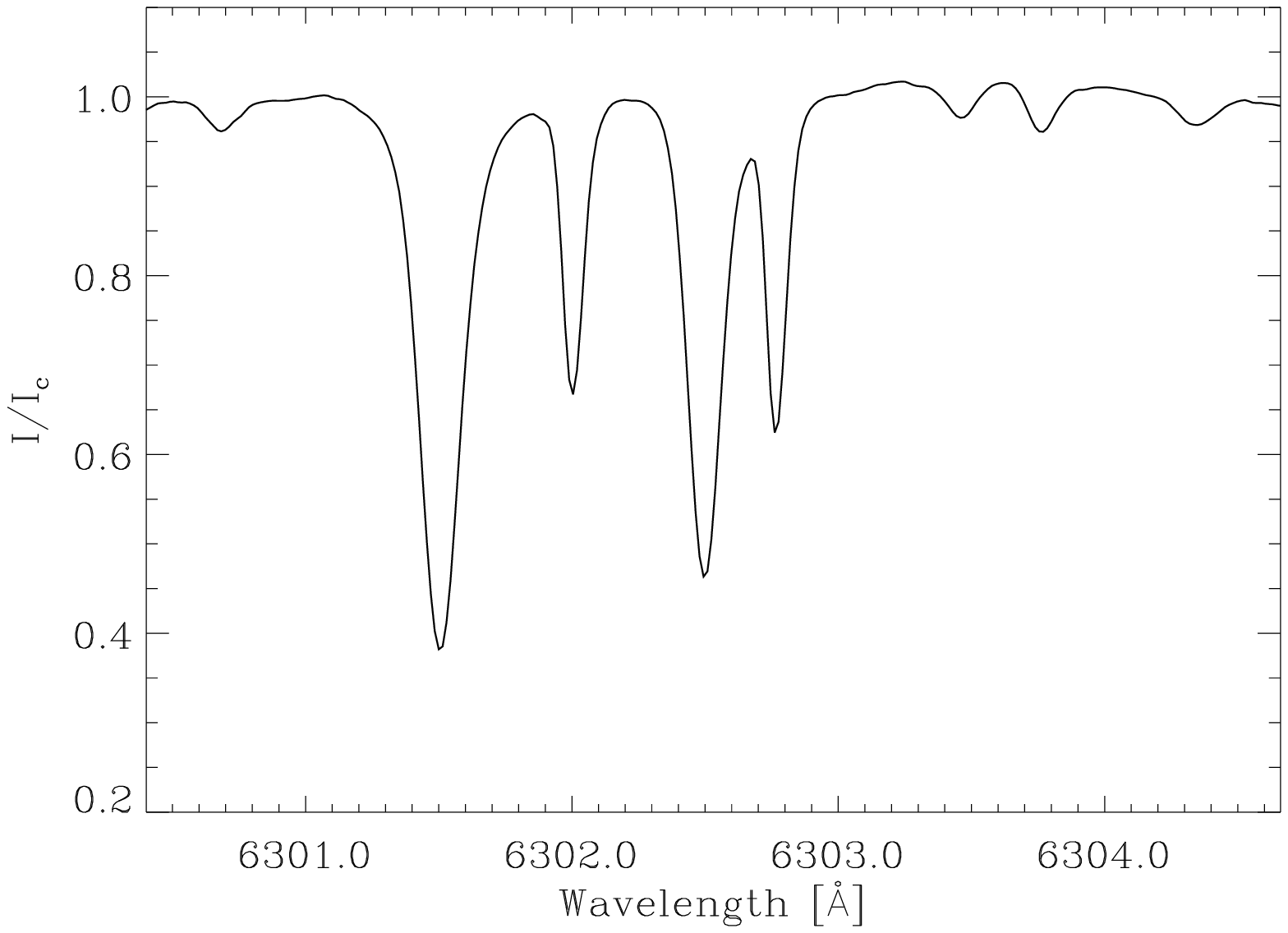}
 \plottwo{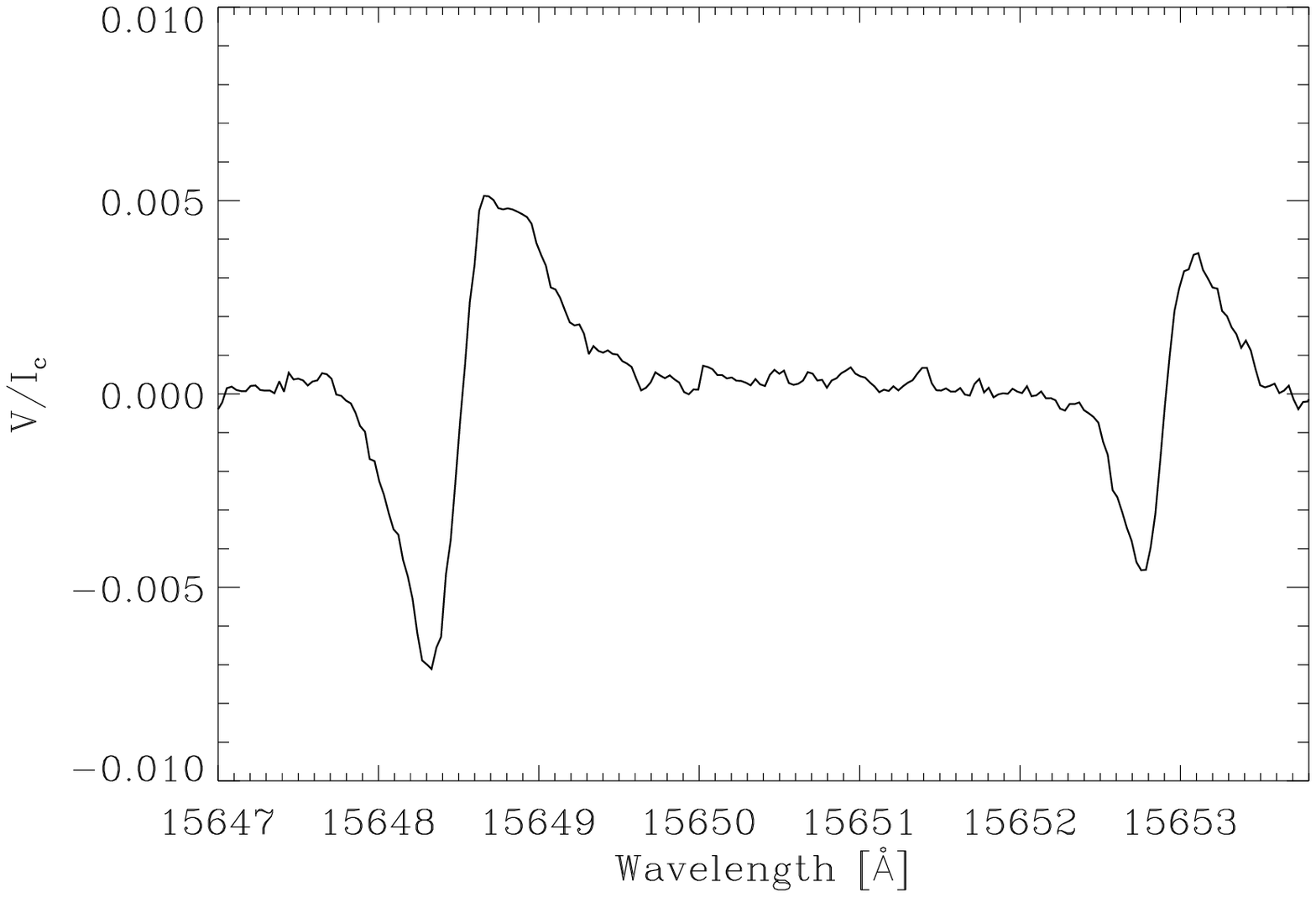}{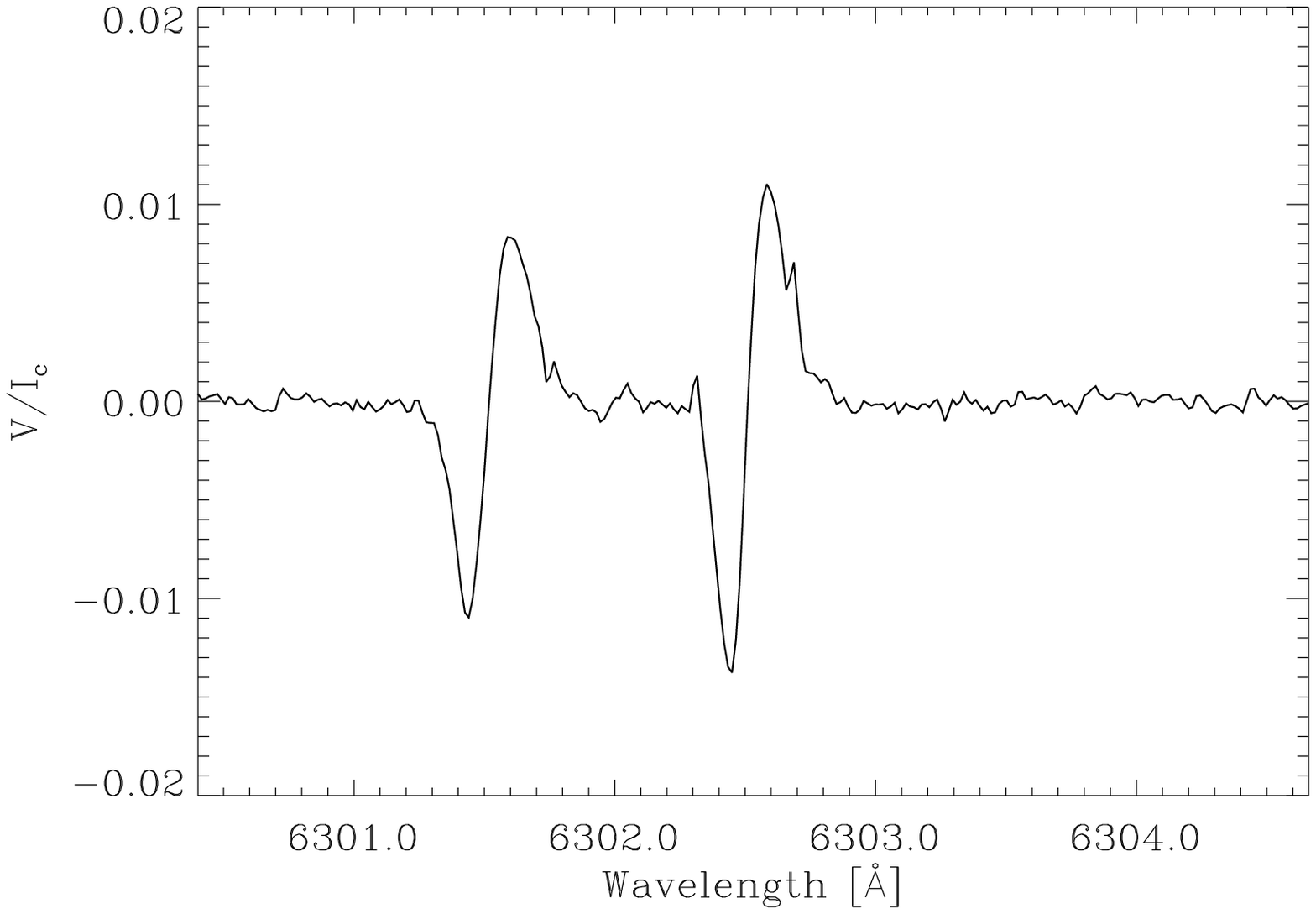}
 \caption{Example of the spectropolarimetric data that we have analyzed in this work. These observations
have been obtained in an internetwork region of the quiet Sun \citep{martinez_gonzalez_spw4_06,martinez_gonzalez06b}.
The upper panel shows the Stokes $I$ profiles in two different spectral regions, one in the near-IR and the other in the
visible. The lower panel shows the circular polarization Stokes $V$ profiles.
\label{fig:spectropol_data}}
\end{figure*}

Except for PCA and AANNs, no other dimension reduction methods have been applied to the field of spectropolarimetry. 
Furthermore, the authors are not aware that any nonlinear dimension
reduction method has been applied to spectropolarimetric
data thus far. In any case, it is always advantageous to have reliable information on the intrinsic dimensionality
of the observed datasets. Although the spatial resolution of solar spectro-polarimetric observations has
improved during the last decades, the resolution elements are typically much larger than the organization
scales in the solar atmosphere. The ensuing mixture of signals inside the resolution element makes it necessary
to use complicated models to explain the observed signals. However, it is fundamental to have
in mind that too complicated models (with a large amount of free parameters) may not be constrained
by the observations.
This paper presents a step forward in the systematic
investigation of observational datasets with the aim of extracting as much information as possible
from the observations. Although we focus on spectroscopic and/or spectropolarimetric datasets, the
philosophy of the approach is applicable to other kinds of data as
well. Nowadays, solar spectroscopic and spectropolarimetric
datasets are becoming very large and some effort is needed to correctly exploit all the information they carry
about the physical processes taking place in the plasma. We review a powerful method presented recently to
estimate the intrinsic dimension of a dataset and we apply it to different observations, analyzing in
detail the consequences. An example of the datasets we are interested here is shown in Fig. \ref{fig:spectropol_data}.
The usual intensity spectrum is shown in the upper panel for two different spectral ranges, while the wavelength 
variation of the circular polarization is shown in the lower panel.

\section{Basic theory}
\subsection{Dimension estimation}
The intrinsic dimension of a dataset is informally defined as the number of parameters
that is needed to describe it. 
In other words, given a dataset consisting of $N$ different observations, each one made of
an $M$-dimensional vector, we seek the dimension $m$ of the nonlinear manifold that captures the
behavior of the $N$ vectors.
As already stated, the dimension of this nonlinear manifold is 
smaller than that of the original space. This is a consequence of the large number of correlations that
are present among the data. Consequently, we can consider that the number of parameters $m$ that 
we need to describe our observations fulfills $m \ll N$, always keeping in mind that these parameters
have to be able to describe the whole nonlinear manifold.

Dimension estimation methods can be classified in two groups. The first group contains all the methods
that rely on the diagonalization of a given correlation matrix (either
linear, such as PCA, or nonlinear, such as
kernel PCA). These methods estimate the dimension by calculating the number of eigenvalues greater than
a given threshold. As discussed above, these methods depend largely on the ability to capture the
nonlinearity of the manifold. Moreover, the estimated dimension
critically depends on the threshold chosen,
a quantity that is often difficult to define and has some degree of arbitrariness. 
However, model complexity information may be incorporated into the
dimension estimate problem 
to generate a less arbitrary threshold \citep{asensio_ramos06}. 

The second group contains 
methods based on geometry, especially important in determining the fractal dimension of dynamical
systems. The analysis of dynamical systems reveals that a large fraction of them
exhibit trajectories in the phase space that have not an integer dimension. After 
\cite{mandelbrot82}, these objects have been named \emph{fractals}. Powerful method have 
been developed to estimate fractal dimensions. Many of them are based on the box-counting
dimension \citep[e.g.,][]{kolmogorov58,hilborn00}. This method estimates the dimension of a given
dataset by calculating the minimum number of ``boxes'' of side $r$ that are needed
to cover the space occupied by the dataset. It is expected that the number of boxes $N(r)$
increases when $r$ decreases, so that the box-counting dimension of the dataset is given by the 
following scaling relation:
\begin{equation}
N(r) = \lim_{r \to 0} k r^{-m},
\end{equation}
where $k$ is a constant. From this, the dimension is obtained by taking logarithms:
\begin{equation}
m = -\lim_{r \to 0} \frac{\log N(r)}{\log r}.
\end{equation}
For the case of simple low dimensionality datasets, it is easy to verify that the box-counting dimension gives
the correct answer. For instance, if our data are distributed on a straight line of length $L$ in a two-dimensional
space, it is easy to demonstrate that $N(r) = L/r$, so that $m=1$. However, this estimation based on
box-counting suffers from computational problems for complex dataset and the computational work grows
exponentially with the dimension of the original data. Another less computationally intensive
dimension estimation method (and probably the most popular thus far) was introduced by \cite{grassberger83} and
employs the correlation dimension. This correlation dimension is based on the observation that
in a $N$-dimensional dataset, the number of pairs of points that are closer than a distance $r$ is
proportional to $r^m$, where $m$ is the correlation dimension. Refinements to this method have 
been introduced recently to overcome some of its limitations \citep{camastra02,kegl02}.

\subsection{Maximum likelihood dimension estimation}
A recent approach to the estimation of dimension has been suggested by \cite{levina_bickel05}.
It has been obtained by applying the principle of maximum likelihood to the nearest neighbor 
distances, resulting in a method for dimension estimation that ourperforms the previous ones.
Let $\mathbf{x}_i$ represent one of the $N$ $M$-dimensional vectors that constitute the observed
dataset. The maximum likelihood dimension estimation assumes that the data points surrounding
$\mathbf{x}_i$ can be correctly described with a uniform probability distribution function. As a 
consequence, the nearest neighbor distances follow a Poisson process. This also leads to 
an easy calculation of the statistical properties of the estimator.
We assume that the observed dataset represents a nonlinear embedding of a lower dimensional
space of dimension $m \ll M$. \cite{levina_bickel05} demonstrated that the maximum likelihood
estimator $\hat m$ of the intrinsic dimension (MLEID) can be written as:
\begin{equation}
 \hat{m}_k(\mathbf{x}_i)^{-1} = \frac{1}{k-2} \sum_{j=1}^{k-1} \log \frac{T_k(\mathbf{x}_i)}{T_j(\mathbf{x}_i)},
 \label{eq:estimation}
\end{equation}
where $T_k(\mathbf{x}_i)$ represents the Euclidean distance between point $\mathbf{x}_i$ and
its $k$-th nearest neighbor. Note that the previous equation is only valid for $k>2$.

The outcome of the previous equation depends critically on the number of neighbors $k$ that are
taken into account. The reason for this is that $k$ sets the scale at which we are analyzing
the dataset, and it is possible that the data have a different dimension at different scales. For instance, this
is the case for a set of points in a two-dimensional space distributed according to a gaussian
density. At very small scale (small value of $k$), we see individual points and the
dimension is close to 0. At larger scales, the dimension reaches the value of 2 \citep[e.g.,][]{kegl02}.
Like other methods, the quality of the estimated dimension usually degrades when $k$ increases as a 
consequence of the finite number of observations in the dataset \citep{levina_bickel05}.

The previous equation is interesting because it allows us to give local estimations of the intrinsic 
dimension, in cases where one expects it to change from point to point. Although more work needs
to be done, in principle it permits to locate points in the dataset that present anomalies
with respect to the average behavior. In any case, it is important to take into account
that large fluctuations can be expected in the estimation of the local dimension and the 
information provided by Eq.~(\ref{eq:estimation}) has to be analyzed with care. However, if we 
assume that the observed dataset
belongs to the same manifold, it is more convenient to use an estimation that takes into 
account all the points in the dataset. \cite{levina_bickel05} propose to use the following
estimation:
\begin{equation} 
  \hat{m}_k = \frac{1}{N} \sum_{i=1}^{N} \hat{m}_k(\mathbf{x}_i),
  \label{eq:average1}
\end{equation}
which is simply an average over the complete dataset. On the contrary, it has been suggested
elsewhere\footnote{http://www.inference.phy.cam.ac.uk/mackay/dimension} that, due to the
mathematical structure of Eq.~(\ref{eq:estimation}), it makes more sense and is more stable to carry out 
the average of the inverse of the estimators:
\begin{equation}   
  \hat{m}_k^{-1} = \frac{1}{N (k-1)} \sum_{i=1}^N \sum_{j=1}^{k-1} \log \frac{T_k(\mathbf{x}_i)}{T_j(\mathbf{x}_i)},
  \label{eq:average2}
\end{equation}
so that the estimation of the dimension is given by $1/\hat{m}_k^{-1}$.
We have verified that both estimates give almost the same value for the dimension, although the latter has
a better behavior for small values of $k$.

The computational cost of this method \citep{levina_bickel05} is mainly dominated by the calculation of the
$k$ nearest neighbors for every point $\mathbf{x}_i$. The computational cost of evaluating Eqs. (\ref{eq:average1}) or
(\ref{eq:average2}) turns out to be almost negligible. Since we are not dealing with too large datasets, our
calculations rely on the calculation of the distances among all the points, so that the computational work is
essentially proportional to $N^2$. However, alternative ways of calculating (exact or approximate) nearest 
neighbors have been developed, the majority of them being based on the construction of efficient tree-like
structures that highly reduce the computational work.

\section{Artificial datasets}

\subsection{Cases with a known number of dimensions}
\begin{figure*}[!ht]
 \plottwo{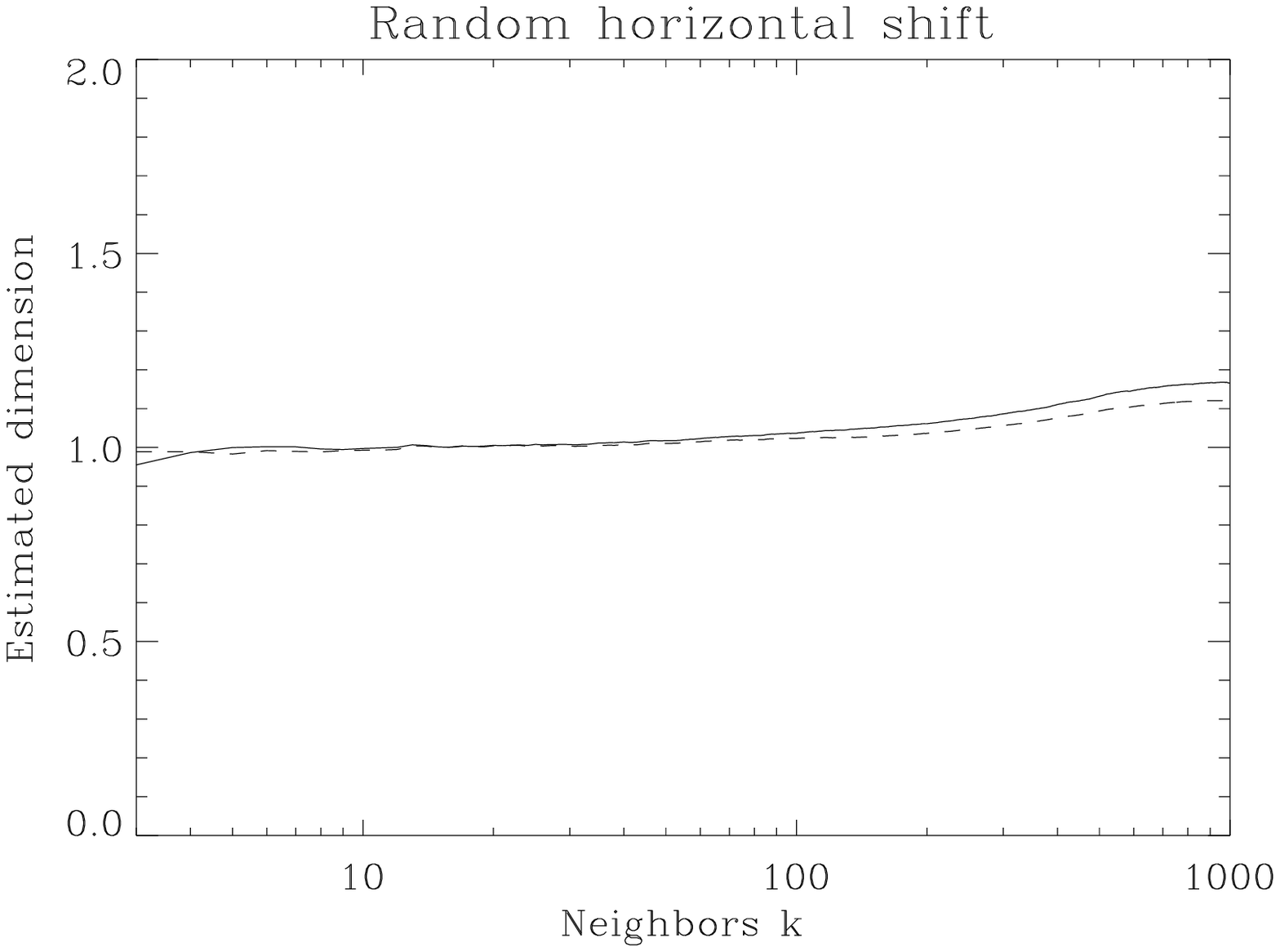}{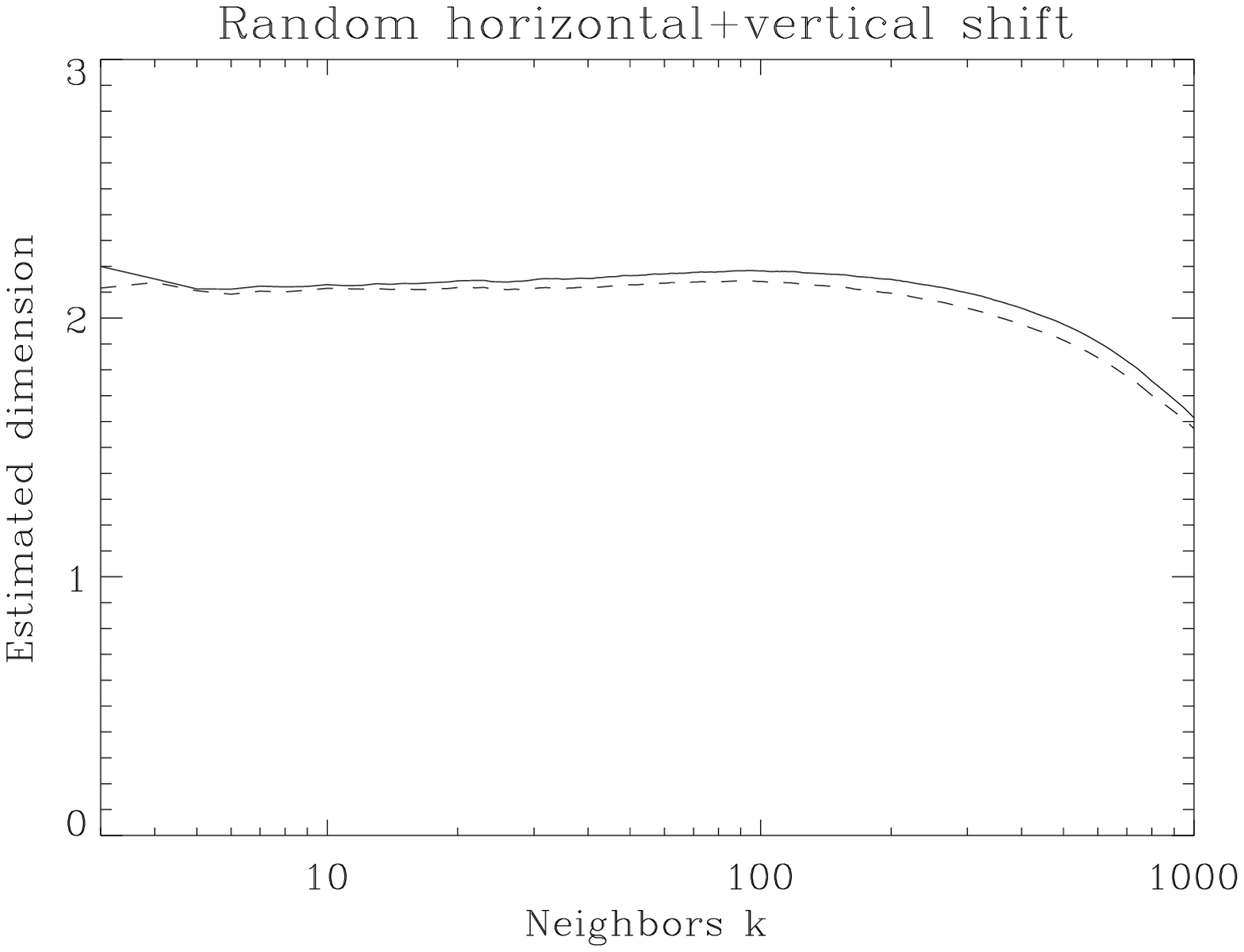}
 \caption{Estimated dimension for two simple cases. In the left panel we show the result when the database
 consists of a single profile that is horizontally shifted by a random sub-pixel quantity. The method correctly
 yields a value of 1 for the dimension. The right panel shows the result when an additional vertical shift is
 applied, giving the correct value of 2.\label{fig:known_cases}}
\end{figure*}

In order to show the reliability of the method introduced by \cite{levina_bickel05}, it is
of interest to test it with datasets of known low dimensionality. 
Although these tests present nothing new with respect to what is already known 
\citep[e.g.,][and references therein]{levina_bickel05}, we consider them necessary to
indicate the potential of these methods.
To this aim, we selected
a particular Stokes~$I$ profile observed with the Tenerife Infrared
Polarimeter
\citep{martinez_pillet99} of an internetwork region of the quiet Sun \citep{martinez_gonzalez_spw4_06}.
With this profile we generate a dataset of 2000 elements by performing
a random horizontal (i.e., in the wavelength direction) shift. The values of the
shift obey a gaussian distribution. 
The estimated dimension is shown in the
left panel of Fig~\ref{fig:known_cases}. Due to the possible variation
of the dimension with the scale at which 
the data are analyzed, we plot the estimated dimension for each value of $k$. When $k$ is small, we 
are referring to small scales while the scale increases when $k$ increases. Because the dataset 
is probably not dense enough to correctly sample the whole nonlinear manifold, there may be
a systematic deviation from the correct dimension for large values of $k$. 
The solid line presents the estimation of the intrinsic dimension obtained from Eq.~(\ref{eq:average1})
while the dashed line presents the estimation given by Eq.~(\ref{eq:average2}). Note that they both
yield similar values for the dimension, which is actually the correct one (since we have allowed only
one degree of freedom). The method has captured the fact that, although these profiles are discretized
in $M=231$ wavelength points, only one parameter suffices to describe the entire dataset.


A further complication is introduced in the artificial dataset by carrying out an additional vertical
shift to the Stokes $I$ profile. The shift follows again a gaussian distribution that is no
correlated with the horizontal shift. The estimated dimension is shown in the right panel of
Fig~\ref{fig:known_cases}. The method correctly gives a dimension of 2. Interestingly, when the vertical 
and horizontal shifts are forced to 
be correlated (for instance, we make them equal) the estimated
dimension is again 1, just as one would expect.

\subsection{Pure noise}
\label{sec:pure_noise}
\begin{figure*}[!ht]
 \plottwo{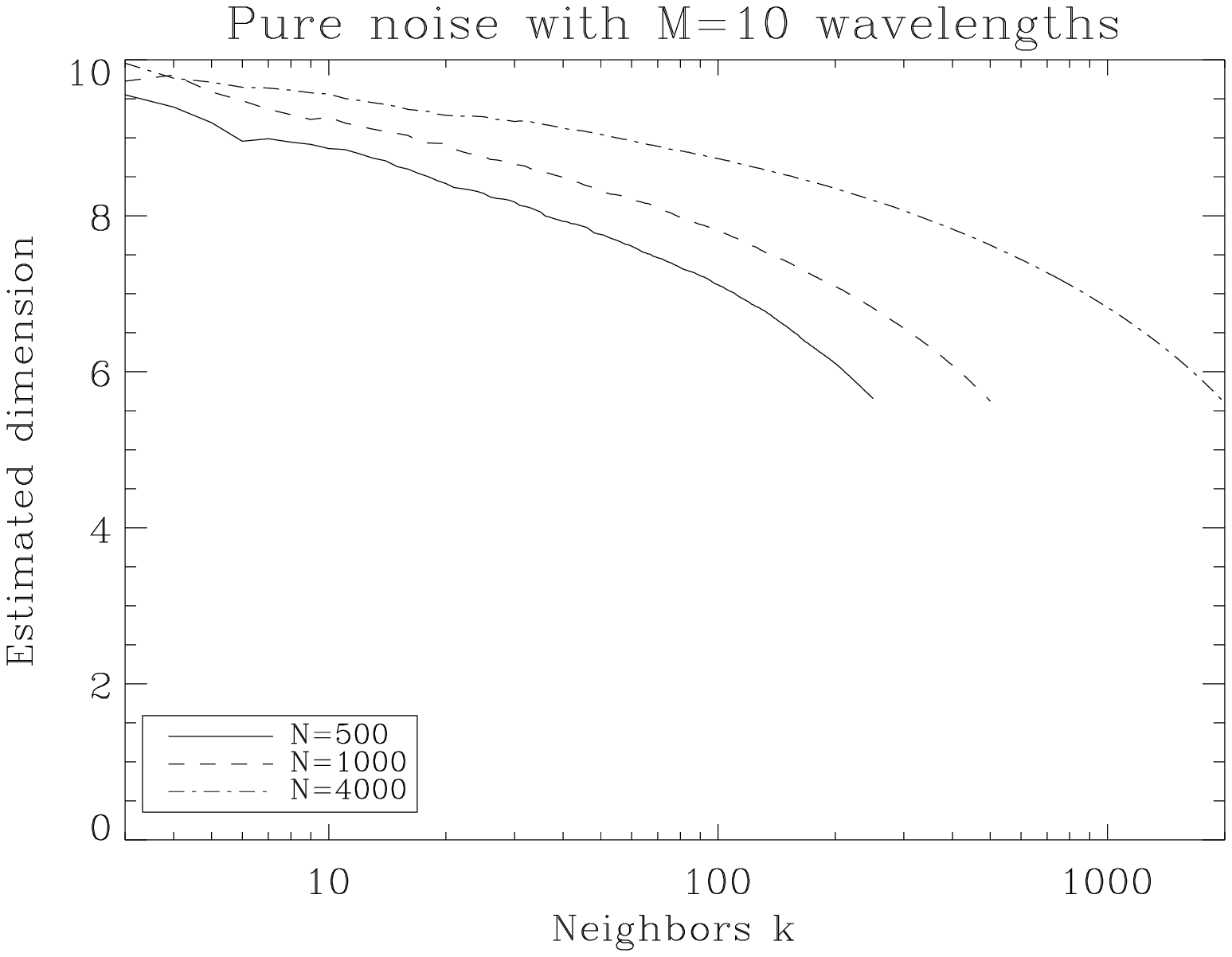}{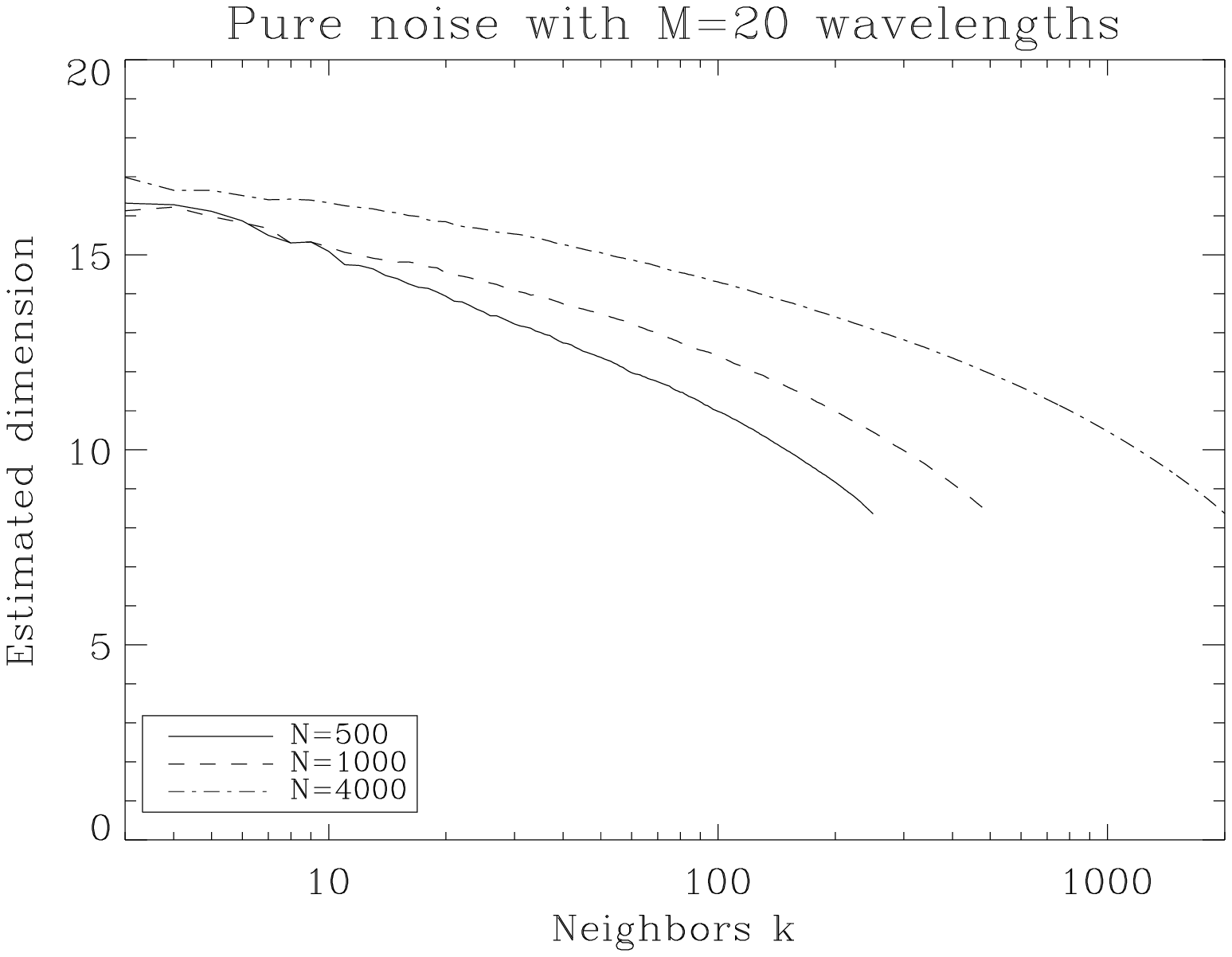}
 \plottwo{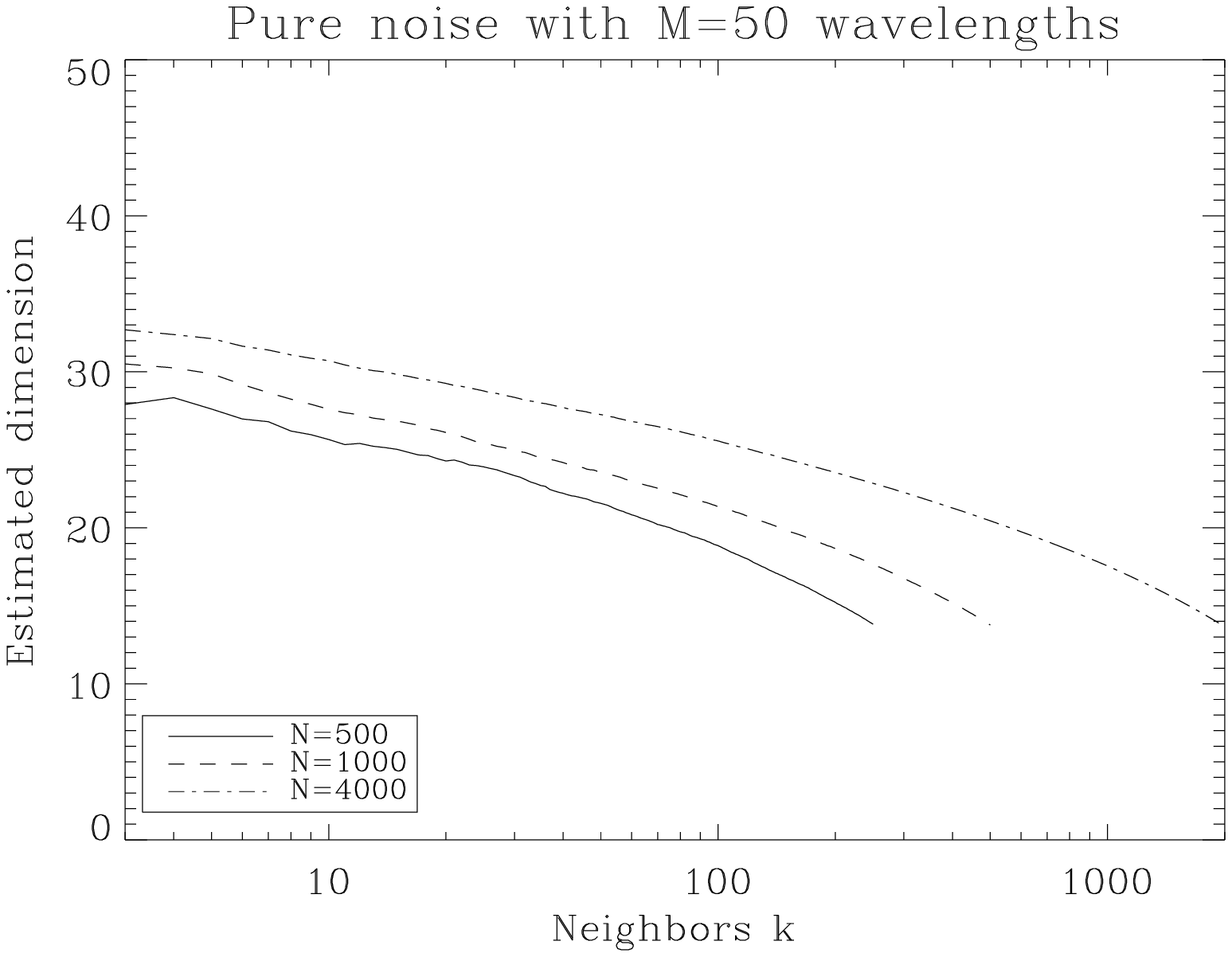}{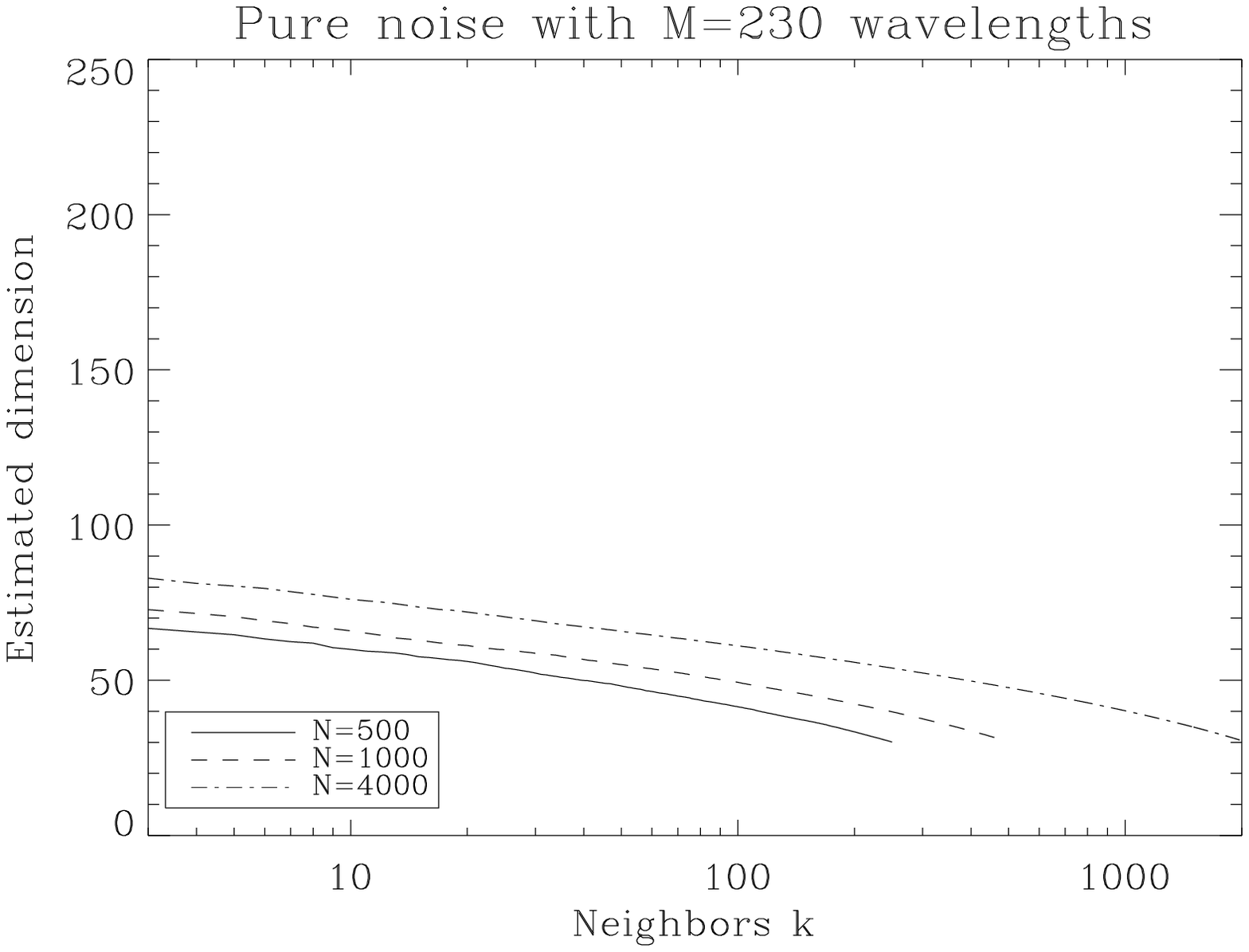}
 \caption{Estimated dimension for profiles composed of noise. The number of wavelength points considered in each
 case is shown in the title.\label{fig:noise}}
\end{figure*}

Noise turns out to be a problem for estimating dimensions. If a dataset is confined inside a manifold
of a high-dimensional space, the inclusion of noise tends to spread the points out of this manifold 
and starts to fill up a larger volume of the original high-dimensional space. Consequently, we
expect that the addition of noise will tend to increase the estimated dimension asymptotically
approaching $M$, the dimension of the original space. We have
generated various sets of profiles with 
different sizes. Each profile consists of a vector of dimension $M$ made of completely uncorrelated
noise following a gaussian distribution. 
The intrinsic dimension of a dataset composed of $M$-dimensional elements of pure noise is equal to $M$ and
we expect the estimators given by Eqs.~(\ref{eq:average1}) and (\ref{eq:average2}) to converge
to this value for sufficiently large values of the number of observables $N$.
The estimated dimensions for each value of $M$ are shown
in Fig~\ref{fig:noise} for datasets of different sizes, from $N=500$ to $N=4000$. In order to
minimize figure cluttering, the curves correspond only to the estimation given by 
Eq.~(\ref{eq:average2}). The same overall pattern is found for Eq.~(\ref{eq:average1}), with a behavior
similar to that found in Fig~\ref{fig:known_cases}. When $M$ is 
small (for instance the case with $M=10$ at the top left
panel), the estimated dimension is very good for small values of $k$. 
It degrades as $k$ increases because the assumption of uniform distribution of the datapoints
breaks for this 10-dimensional space with such a small
number of points. Consequently, the assumptions under which the formalism of \cite{levina_bickel05}
has been developed are not fulfilled and it cannot be applied.
However, it is surprising that it is possible to have a rough estimate with 
a dataset of only $N=500$ elements. When the number of elements of the dataset increases, the curves
asymptotically tend to $M$. For increasing values of $M$, the dimension estimate
is biased towards smaller values, although it is clear that it still yields a reasonable approximation
to the correct value even for very small datasets. Figure~\ref{fig:noise} shows in detail how
increasing the number of elements in the dataset leads to an improved estimation of the 
intrinsic dimension. In the limiting case of a space with extremely large dimension ($M=230$),
the method underestimates the dimension by a factor of $\sim$3.

\subsection{\ion{Fe}{1} database}
One of the fastest techniques for Stokes profiles inversion is based
on a look-up algorithm with PCA coefficients \citep{rees_PCA00}.
Once a model atmosphere (with a given number of parameters) is
selected, a database of models and emerging profiles is 
generated. The database has to be able to correctly sample the space spanned by all the
parameters. Due to computational limitations, the PCA inversion technique has only been applied to the
simple Milne-Eddington atmosphere thus far. The eigenvectors of the
PCA decomposition are then saved, along
with the projection of each element of the database on these eigenvectors, and the Milne-Eddington 
parameters associated with each one. In the inversion process, an
observed set of Stokes profiles is 
projected on the eigenvectors and the corresponding projections are compared to those
saved in the database. Here we have used the PCA database as our
observed dataset. We are interested
in estimating the dimension of the manifold in which these observation ``live''.
In principle, each profile contains 180 wavelength points and the phase space would have
dimension 180. However, correlations between many of these wavelength points (for instance,
all the continuum points that always present the same value) drastically reduce the dimension of
the manifold.

\begin{figure*}
 \includegraphics[width=0.31\hsize,clip]{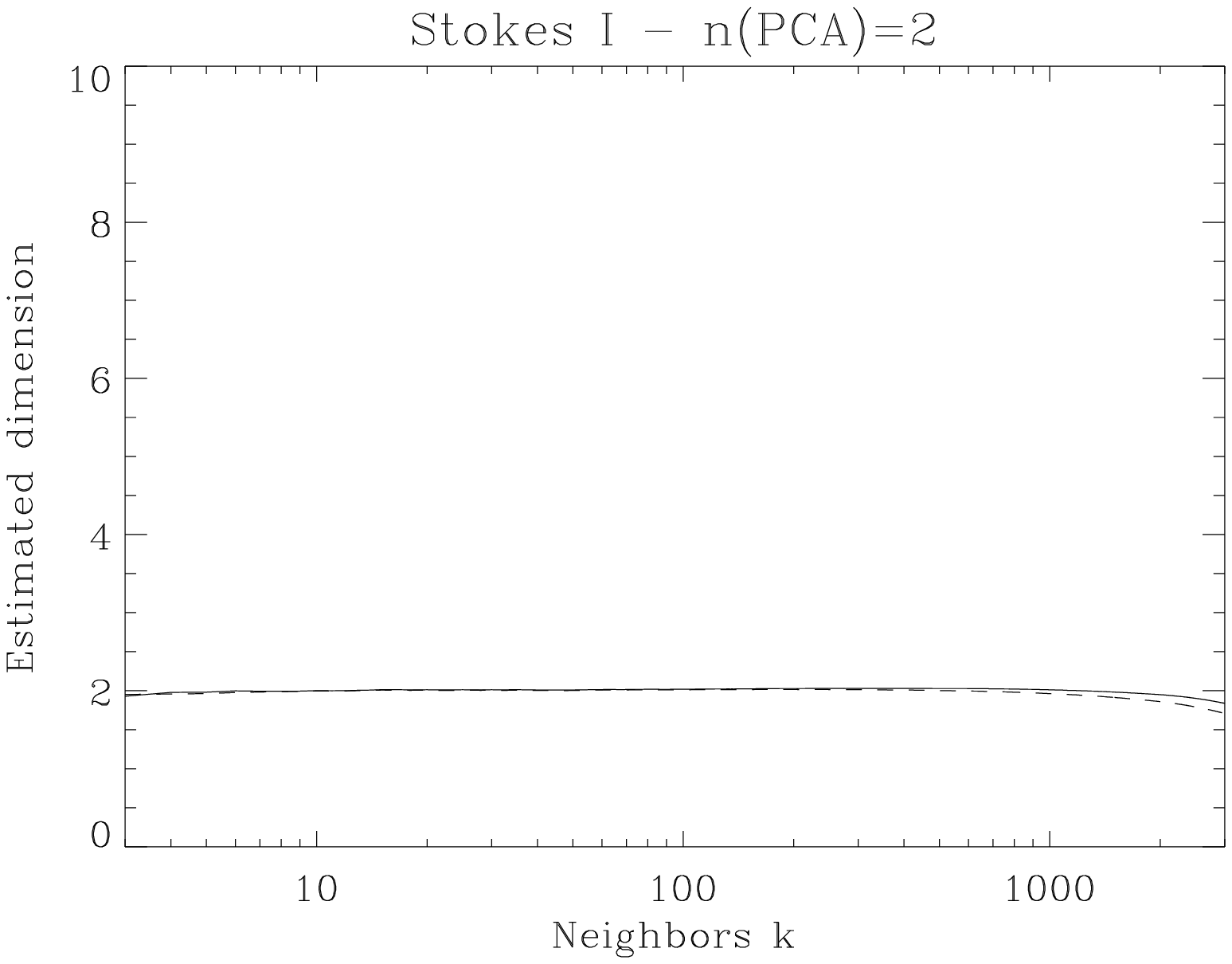}%
 \includegraphics[width=0.31\hsize,clip]{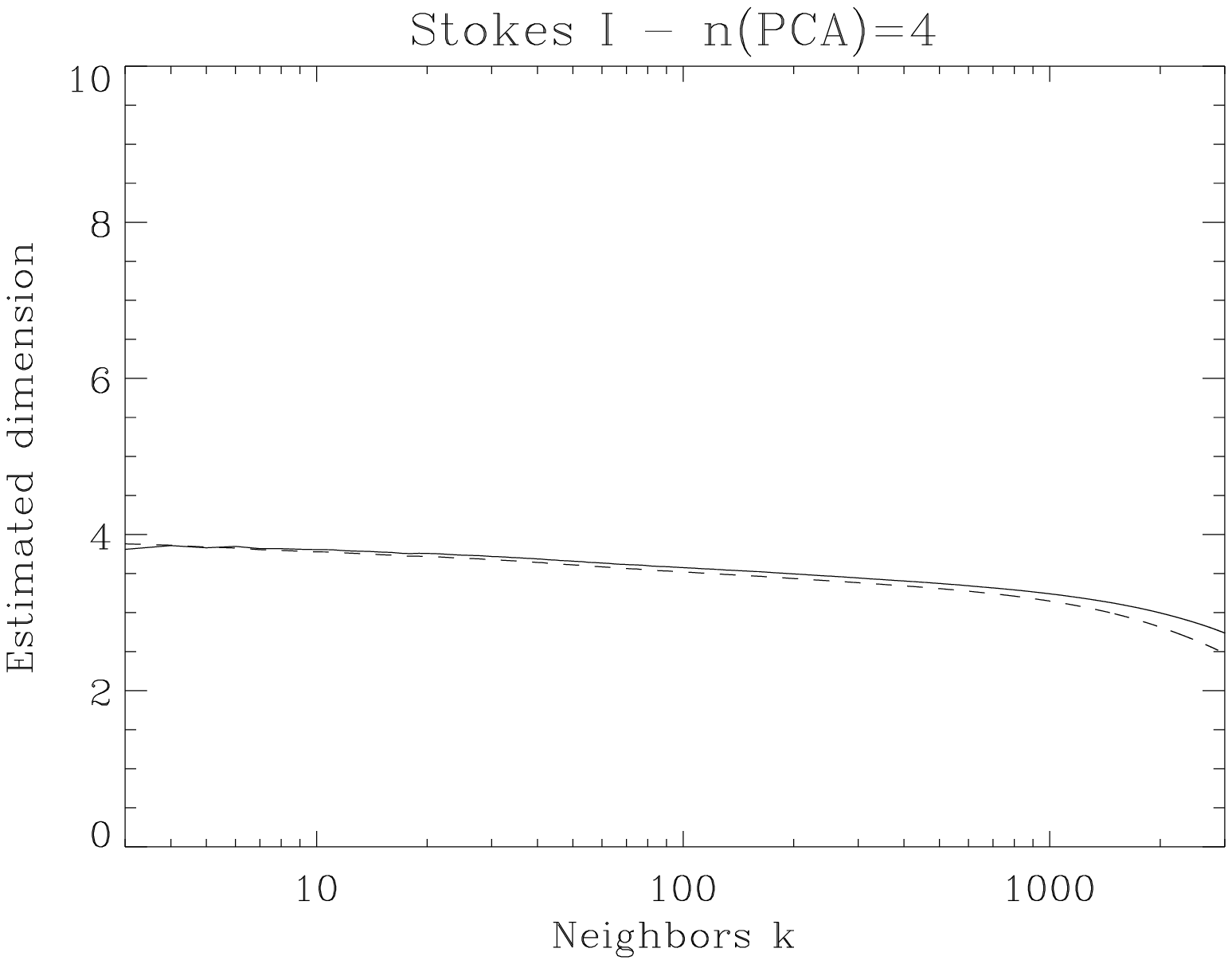}%
 \includegraphics[width=0.31\hsize,clip]{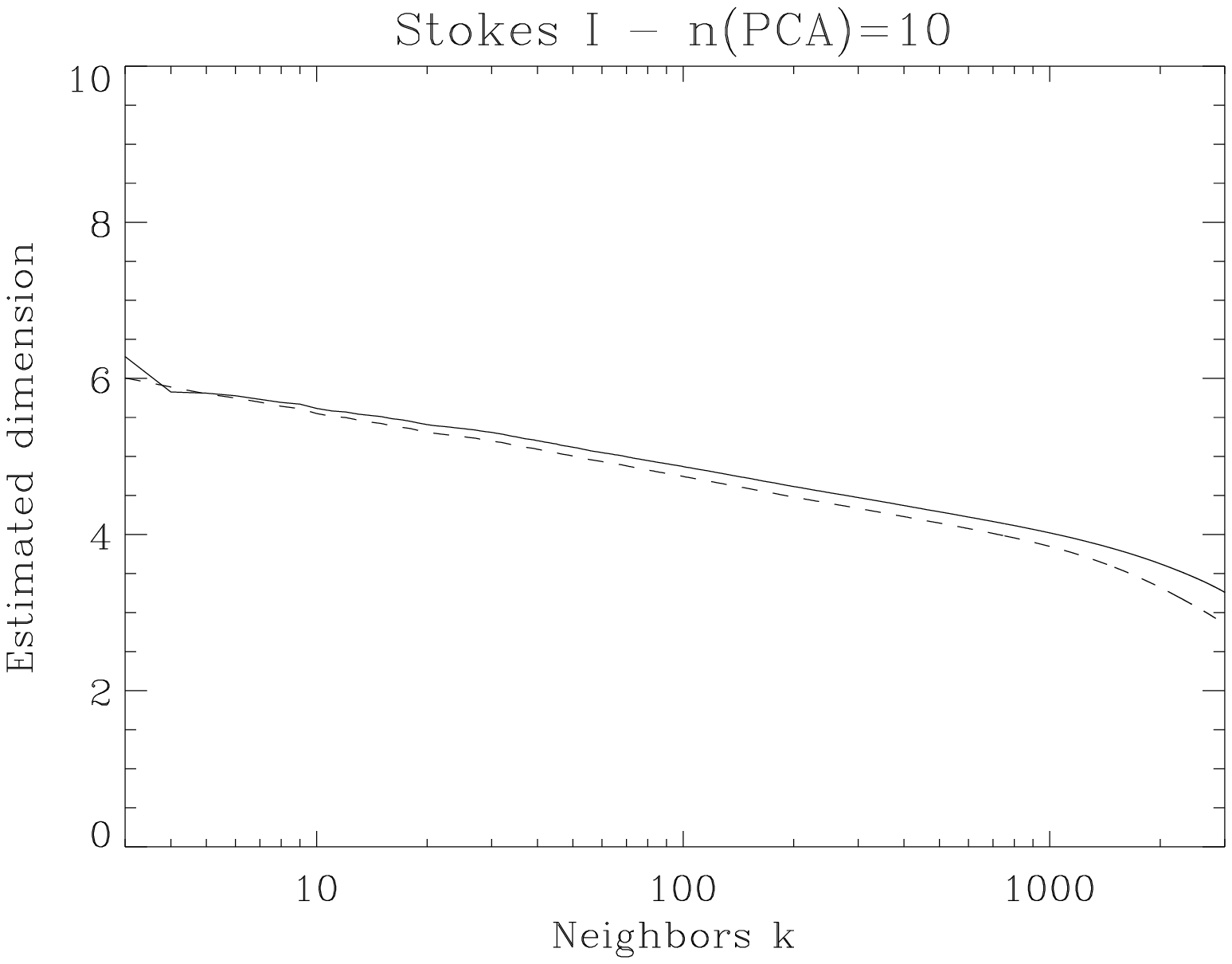}
 \includegraphics[width=0.31\hsize,clip]{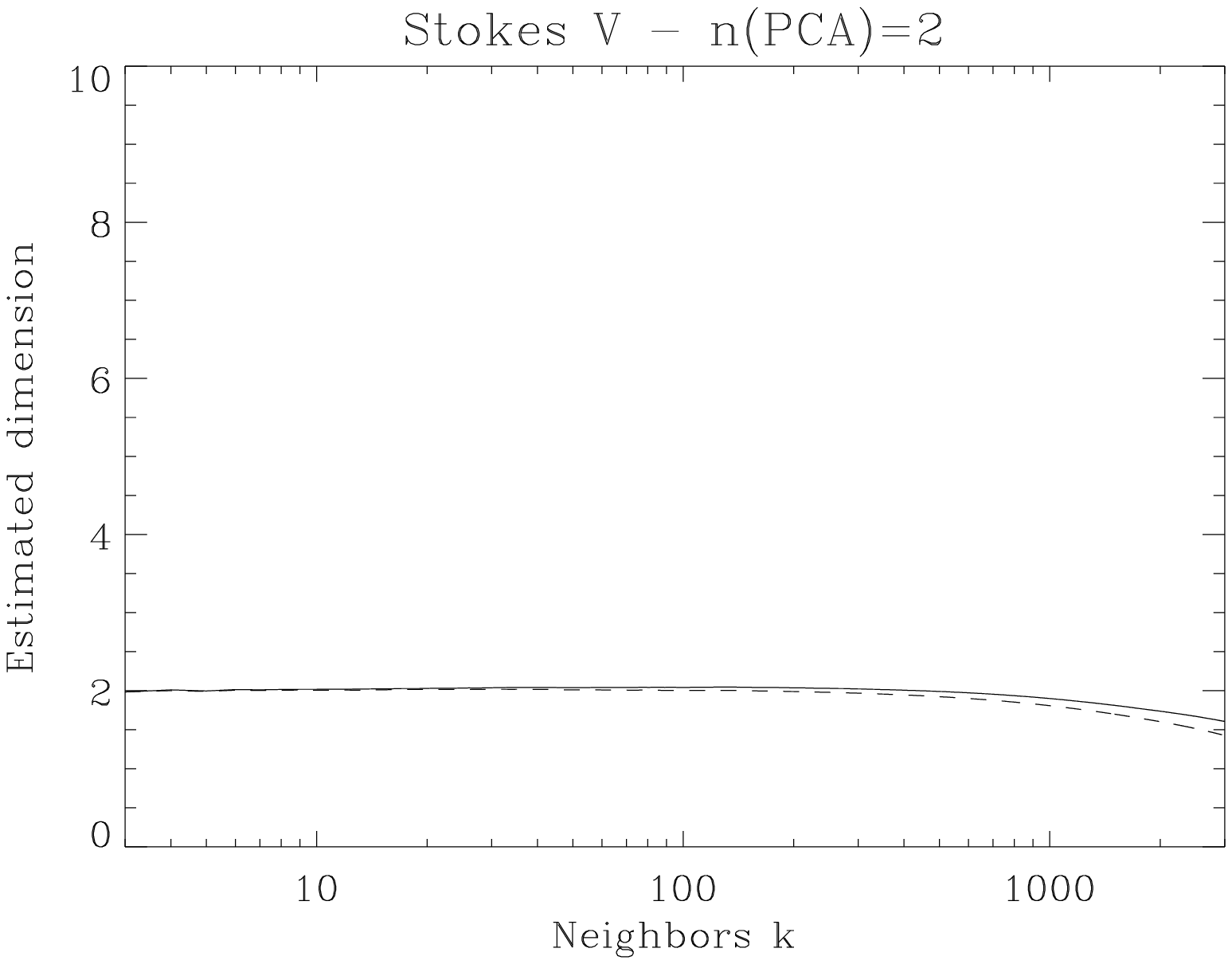}%
 \includegraphics[width=0.31\hsize,clip]{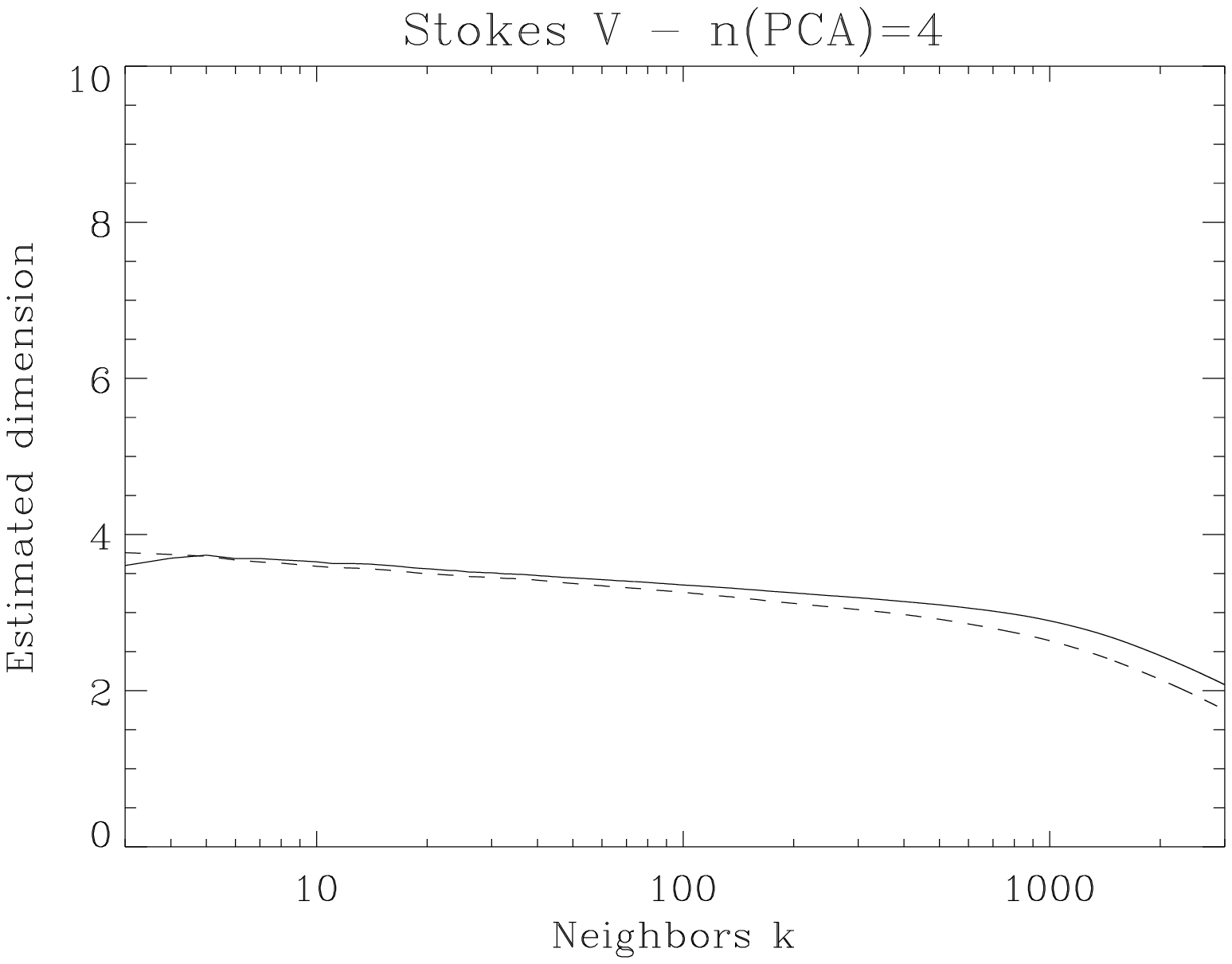}%
 \includegraphics[width=0.31\hsize,clip]{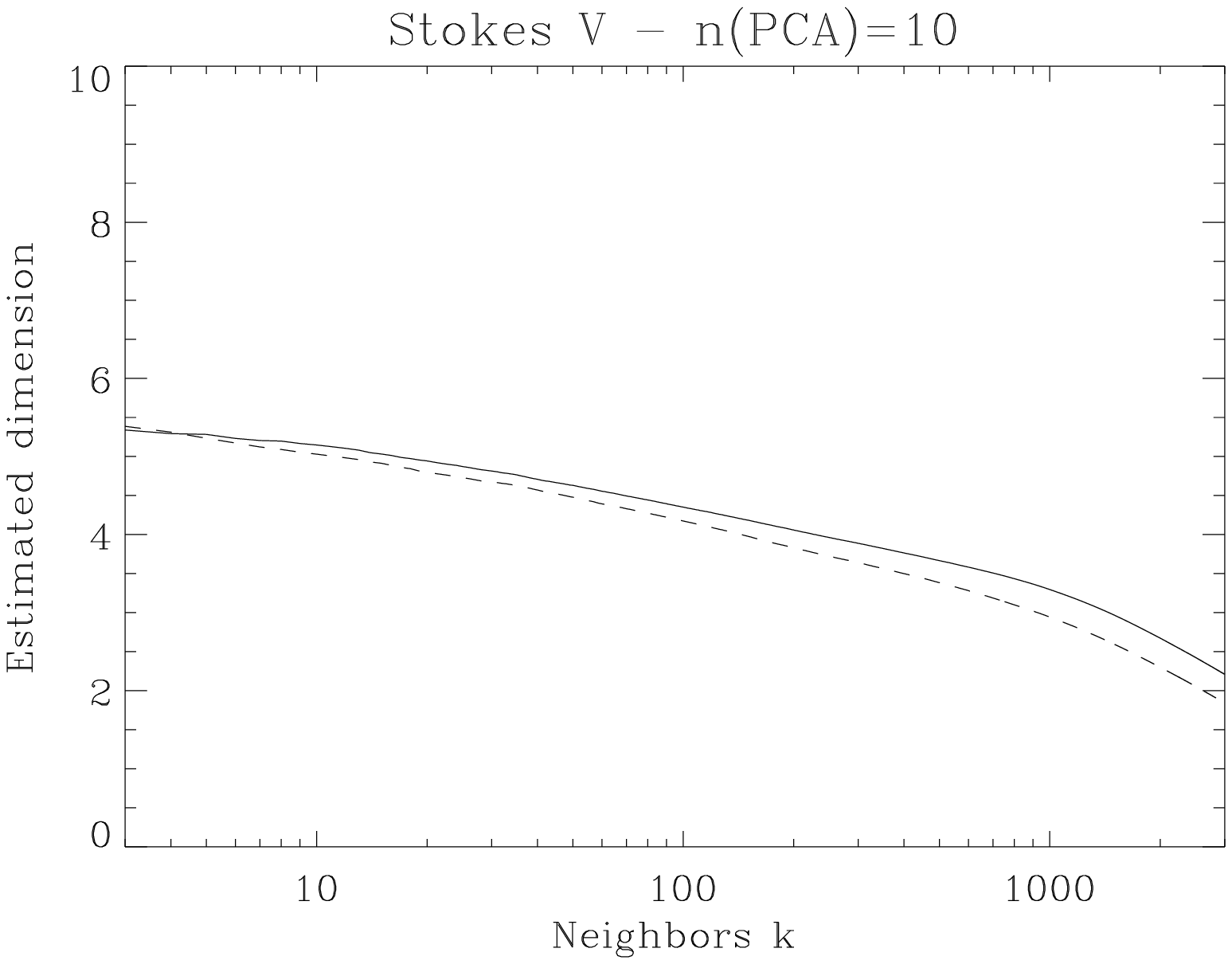} 
 \caption{Estimated dimension for the Stokes I and V database of \ion{Fe}{1}
 as different numbers of PCA components are used
 in the reconstruction.\label{fig:database_pca}}
\end{figure*}

The database that we use consists of $\sim$6200 solar Stokes profiles of the 6301-6302~\AA\ region, where
two \ion{Fe}{1} lines and two telluric lines are visible. Fig~\ref{fig:database_pca} shows the
estimated dimension for the Stokes~$I$ (upper panel) and the Stokes~$V$ profiles (lower panel).
The database is reconstructed from the PCA eigenvectors and the projections of each element of the database
on these eigenvectors. In order to see the information carried out by the eigenvectors, we
show in Fig~\ref{fig:database_pca} the estimated dimension using an increasing number of eigenvectors
$N_\mathrm{PCA}$ in the reconstruction. The trend obtained is very instructive, showing that the 
estimated dimension increases with $N_\mathrm{PCA}$ until a saturation
is reached. The case
$N_\mathrm{PCA}=2$ demonstrates that the first two eigenvectors contain a large amount of information
and they may be seen, as shown by \cite{skumanich02}, as directly
associated with physical parameters. The situation
remains unchanged when reconstructing with $N_\mathrm{PCA}=4$ eigenvectors, while a saturation
is reached when reconstructing with $N_\mathrm{PCA}=10$. This means that, although the number of 
Milne-Eddington parameters defining each element of the dataset is 9,
only 6 are actually needed to 
describe the entire dataset. This is an alternative way of showing the strong degeneracy present in the
6301-6302~\AA\ \ion{Fe}{1} lines \citep{martinez_gonzalez_spw4_06,martinez_gonzalez06}.
Although PCA cannot capture the possible nonlinearity of the 6-dimensional manifold, it can be
shown that the first 6 eigenvectors are sufficient to describe all the elements of the database
with a very small error.

\subsection{The effect of noise}
\begin{figure*}[!ht]
 \includegraphics[width=0.31\hsize,clip]{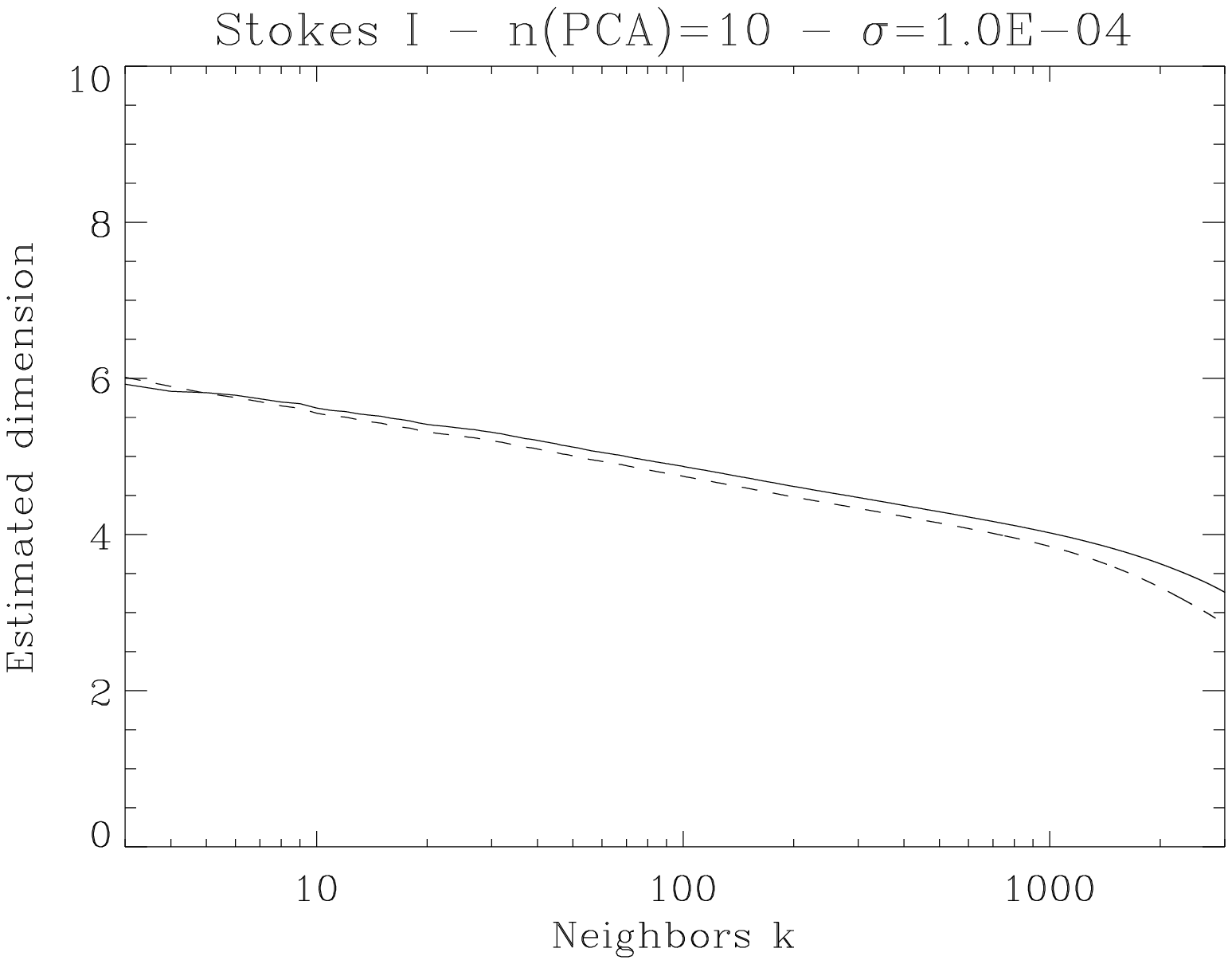}%
 \includegraphics[width=0.31\hsize,clip]{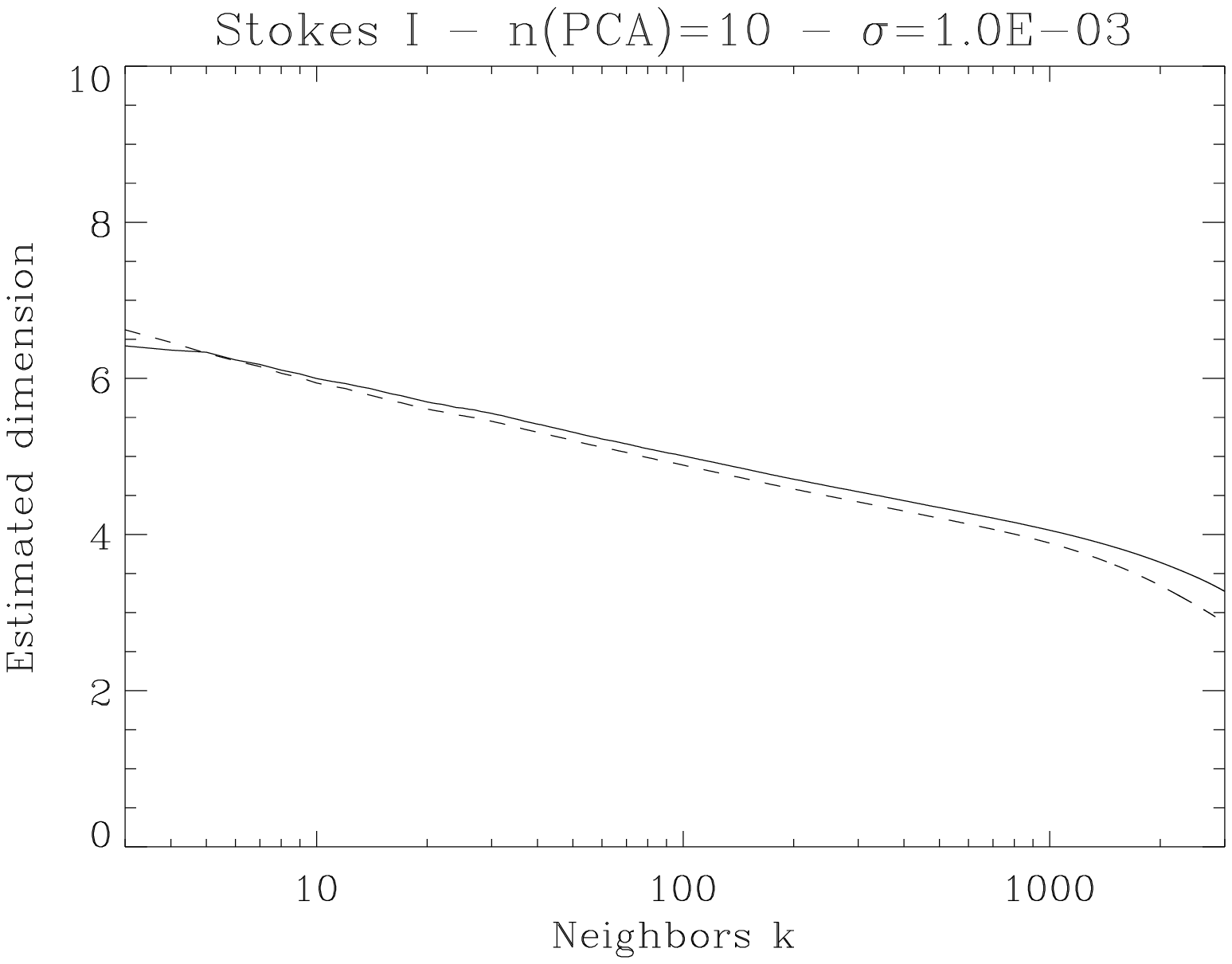}%
 \includegraphics[width=0.31\hsize,clip]{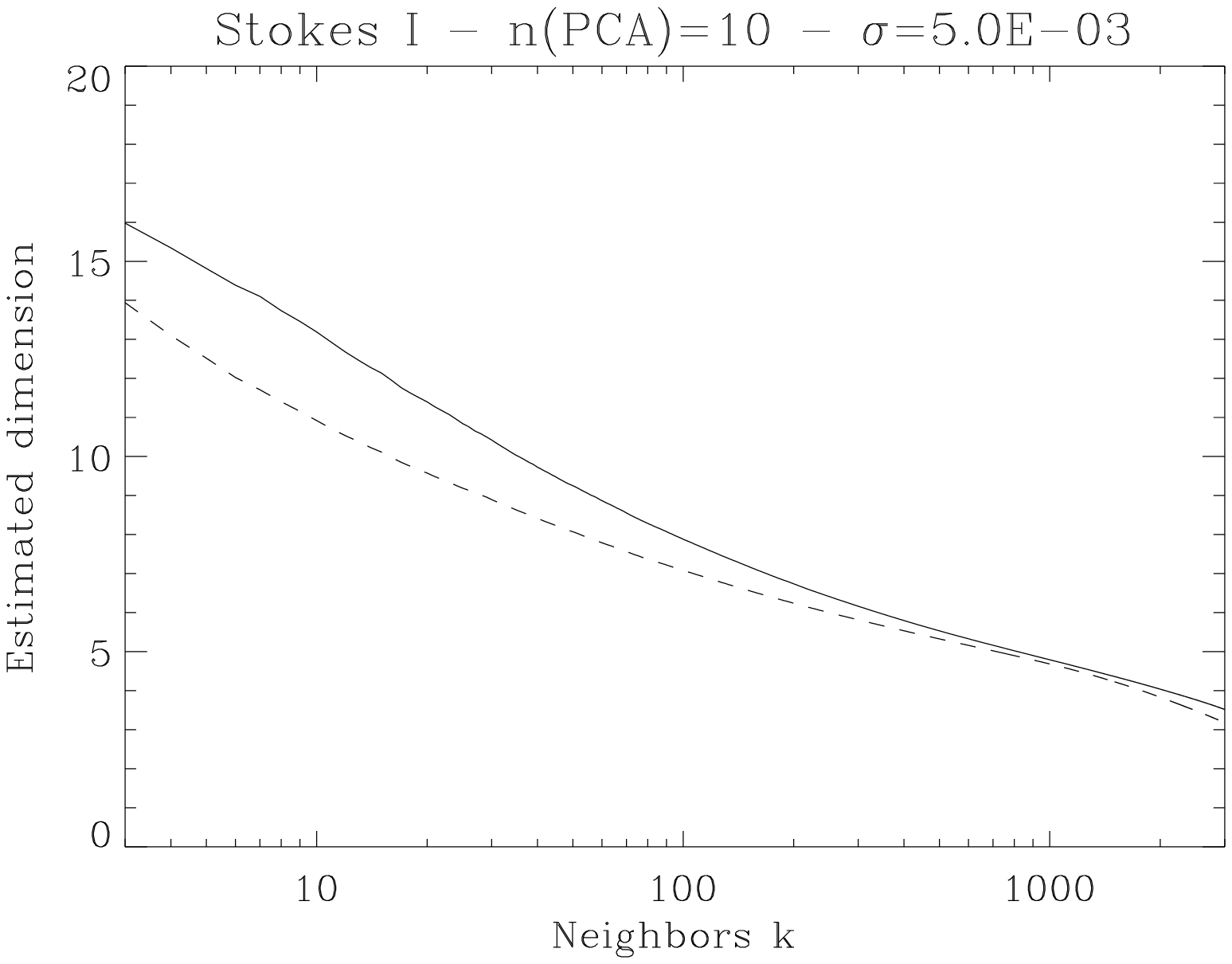}
 \includegraphics[width=0.31\hsize,clip]{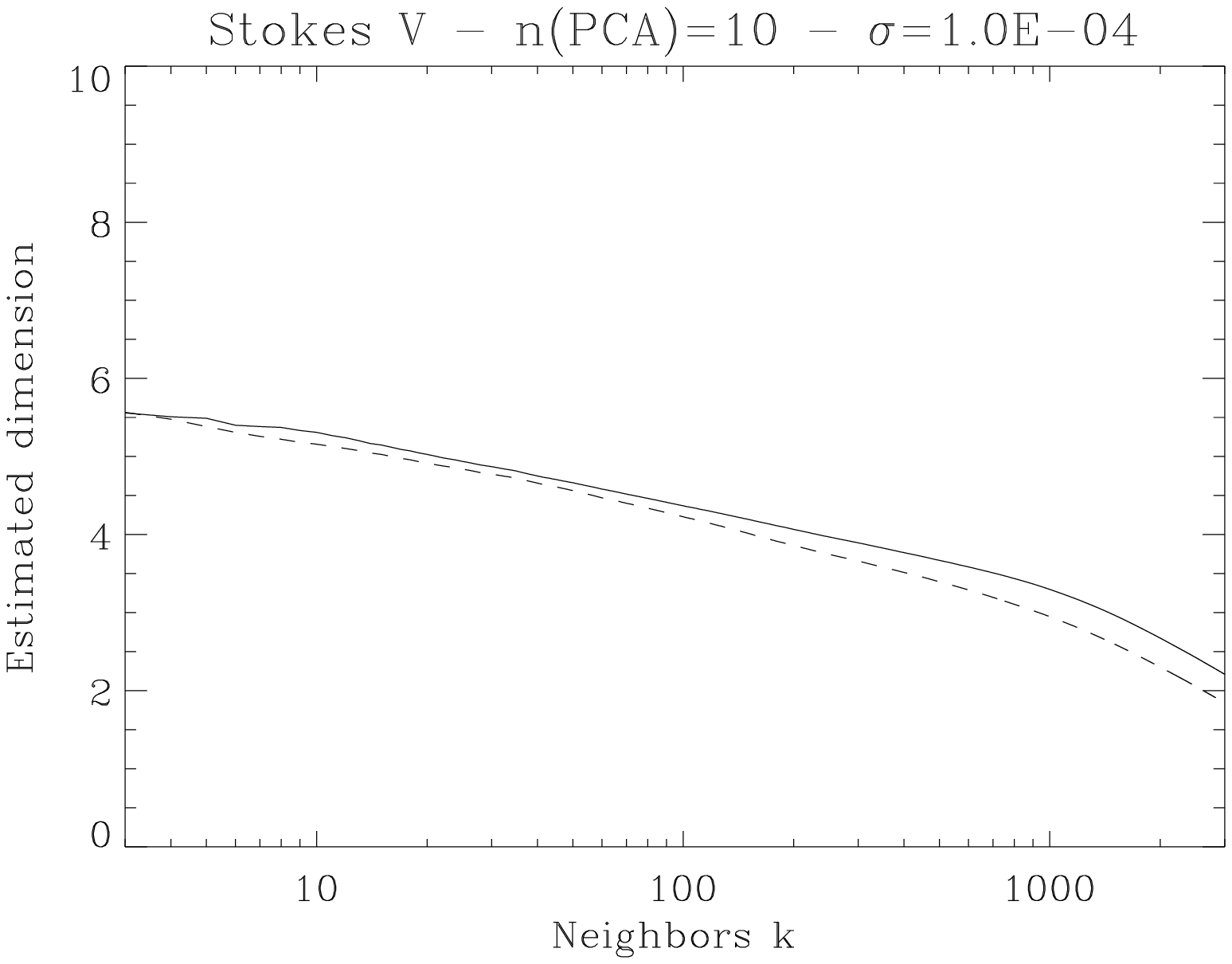}%
 \includegraphics[width=0.31\hsize,clip]{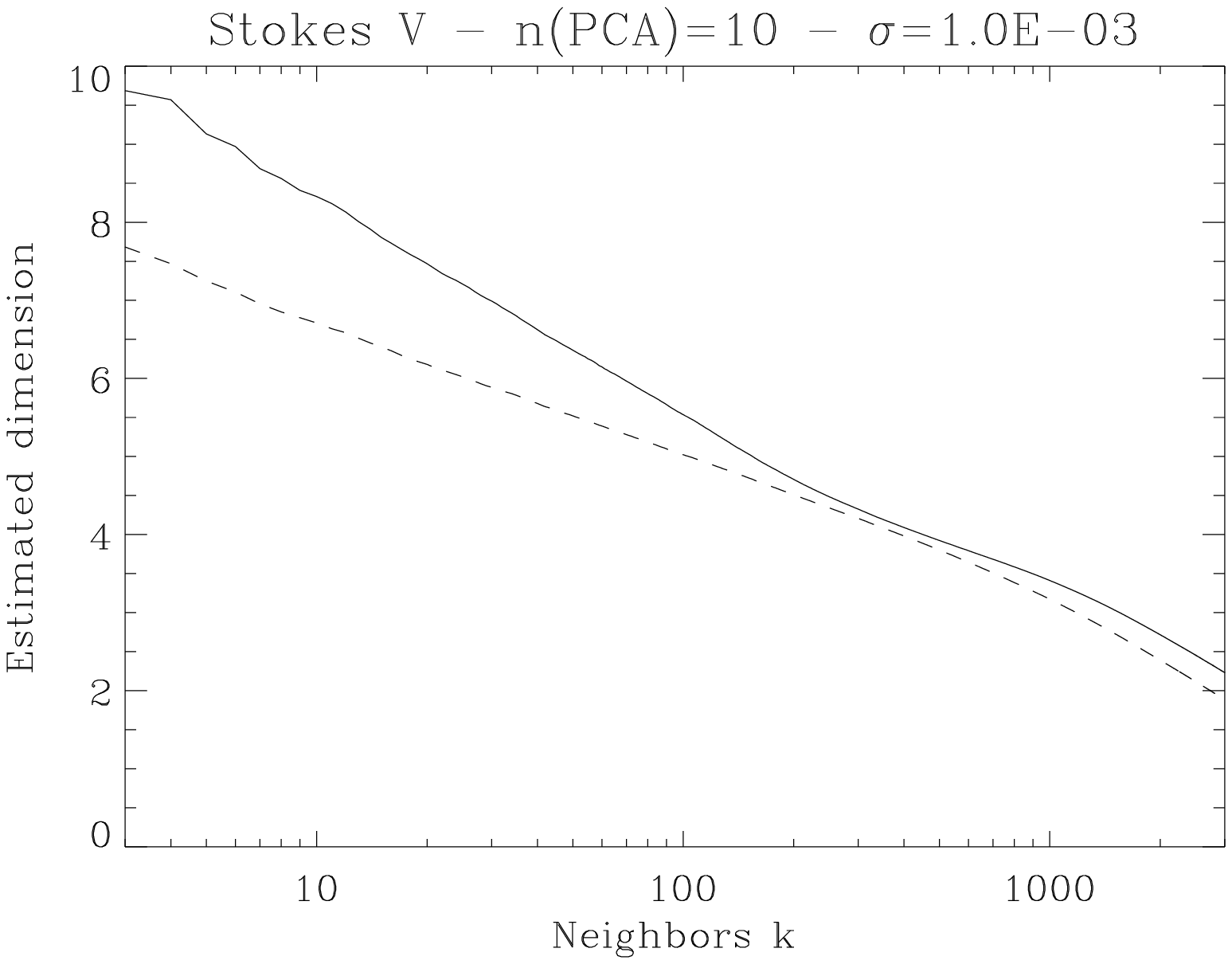}%
 \includegraphics[width=0.31\hsize,clip]{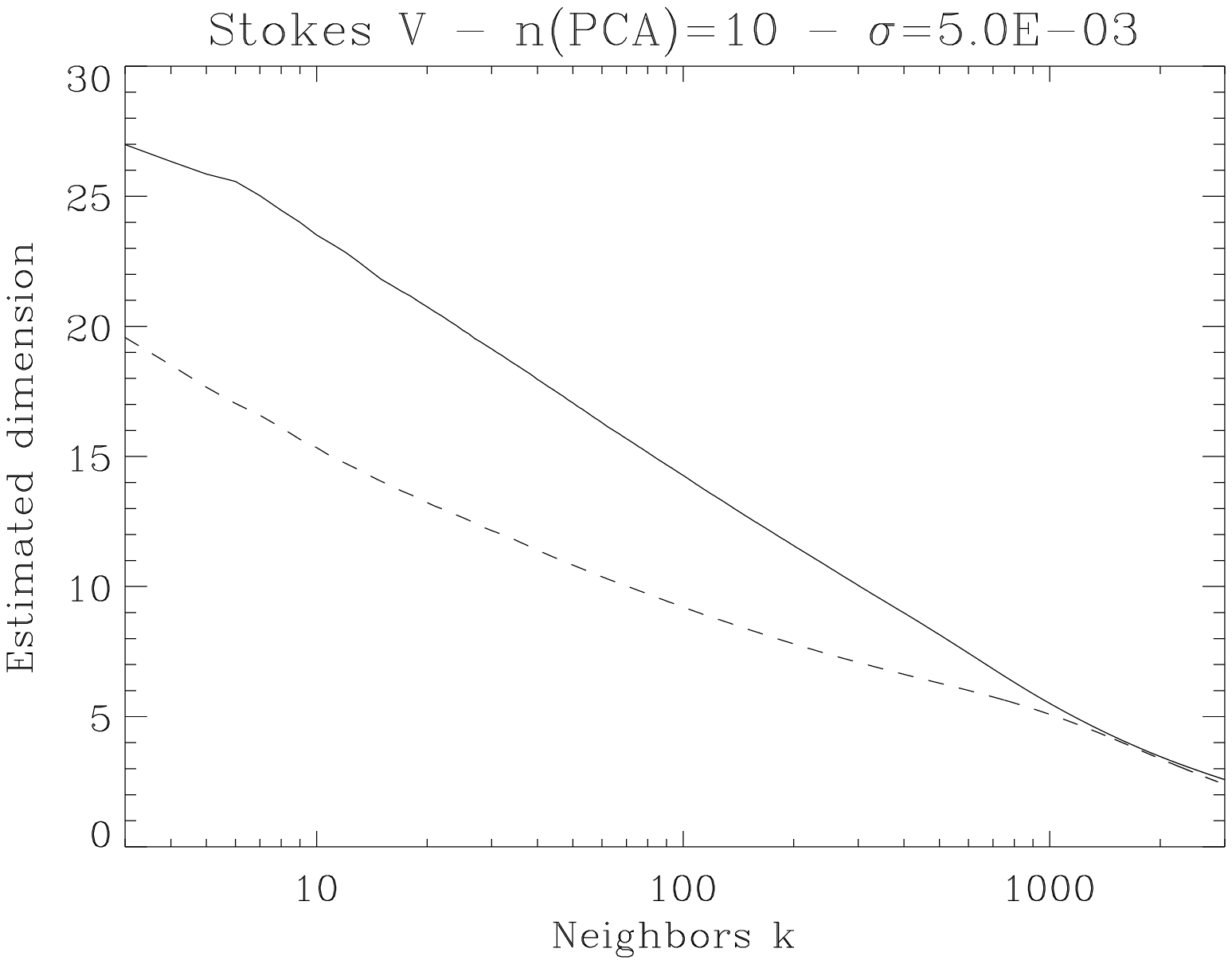} 
 \caption{Effect of noise on the estimated dimension of the synthetic \ion{Fe}{1} dataset. The
 dimension increases with increasing noise, with Stokes~$V$ more sensitive than Stokes~$I$
 due to the difference in amplitude. The noise amplitude $\sigma$ is given in terms of the continuum intensity.\label{fig:noise_effect}}
\end{figure*}

We have shown that the method developed by \cite{levina_bickel05} correctly captures the dimension of
pure noise data. It is even more
important to see how the method behaves when data is corrupted with noise. It is expected that, since the noise reduces
the correlation between some of the components of the $M$-dimensional vectors that represent
the dataset, the estimated dimension will grow when the signal-to-noise ratio decreases. 
A fundamental problem arises because it is very difficult to recognize truly high-dimensional data from 
low signal-to-noise data.
This test has been carried out with the \ion{Fe}{1} dataset reconstructed using all the information
available (we used the first 10 eigenprofiles). We have calculated the estimated dimension
for four different noise levels, given in terms of the standard deviation $\sigma$ of the
gaussian noise in units of the continuum intensity. Since
the typical Stokes~$V$ signals are around 1-2 orders of magnitude smaller than 
Stokes~$I$, a noise of the same $\sigma$ implies a much smaller signal-to-noise ratio
for Stokes~$V$ than for Stokes~$I$. Consequently, we expect the dimension increase to start at
smaller values of $\sigma$ for Stokes~$V$ than for Stokes~$I$. Figure~\ref{fig:noise_effect}
presents the results for three different values of the noise. The value of $\sigma=10^{-4}$
is small enough so that no appreciable difference is found in either
Stokes~$I$ or~$V$ in
the estimated dimension with respect to the case with no noise. When the noise increases to 
$\sigma=10^{-3}$, the Stokes~$I$ profiles still maintain the original dimension while the
estimated dimension for the Stokes~$V$ dataset increases rapidly. It is interesting to note
that the estimated dimension increases faster for small values of $k$. This is 
because the noise is small enough to produce perturbations (cancellation of correlation between 
the $M$ components of each Stokes profile) at very small scales, while the large scale dimension
still remains unchanged. When the noise is increased further, a drastic increase of the dimension
is observed in Stokes~$V$ and a smaller one for Stokes~$I$. Note that for even smaller signal-to-noise
ratios, the estimated dimensions for large values of $k$ would also increase until reaching
(in the limiting case of an infinitely large dataset) a flat dimension
estimate, constant for all scales, and equal to $M$. The typical noise in spectro-polarimetric observations is usually well
below 10$^{-3}$, so that it is apparent from Fig~\ref{fig:noise_effect} that noise is 
not expected to change appreciably the dimensionality of noiseless data.

A consequence of the previous analysis is a possible technique to recognize when data 
is affected in an important manner by noise. Our datasets usually present a large value of $M$ so 
that, in the case of very
large noise, the dimension has to grow until reaching a very large value. A clear effect of
the noise, as stated in section \ref{sec:pure_noise}, is that the estimated dimension rapidly 
increases at small values of $k$
while being held constant for large values of $k$. Thus, if a calculation shows a
estimated dimension that exhibits large values at small $k$ and a steep logarithmic fall 
for large $k$, noise is likely rather important. A caveat is mandatory. This test relies on the
behavior of the MLEID for large values of $k$. As
already pointed out by \cite{levina_bickel05} and also shown here, a degradation of the dimension estimation
occurs for large values of $k$ and the maximum likelihood estimation does not hold. For this reason, 
one has to be cautious with low signal-to-noise ratio observations. Recently, \cite{levina06}
have addressed the problem of dimension estimation of high-noise observations when using the MLEID. Their approach to 
the problem is based on a smoothing of the original dataset, so that the performance of the method is 
greatly enhanced. They find that the estimated dimension for high-noise spectroscopic observations of chemical
mixtures turns out to be extremely large. However, when a certain amount of smoothing is introduced, the
MLEID turns out to be a very accurate estimation of the dimension. Thanks to the low noise in our observations,
our estimations of the intrinsic dimension are surely not dominated by noise and we consider that smoothing is
not necessary.

\section{Observed datasets}
We have shown how the MLEID developed by
\cite{levina_bickel05} works for synthetic data. In this section we focus on real spectropolarimetric
observations. Our aim is to learn about the intrinsic information content of the
data. This may help understand how much complexity can be introduced in the
models used to interpret the observational data. A proposed physical model usually consists of
a set of free parameters that we have to constrain with the observations. It is crucial
to have as much information as possible in the observed dataset so that one can
constrain the model parameters. Obviously, it is undesirable to use too complex models to
infer physical information from a dataset if the observables contain
only a small amount of information. The
parameters used in the physical models are typically non-orthogonal and they usually present
degeneracies because the same observable can be obtained with different sets (finite or infinite) 
of model parameters. Although it is not straightforward to estimate from the intrinsic dimension how many 
parameters one can introduce in the modeling, it obviously should not be much larger than the estimated 
intrinsic dimension. If this number is made much larger, many of these parameters may not be 
constrained by the observations, thus leading to unphysical results or
ill-conditioned inversions.

An important application of the estimated dimension tools we have presented here is to make relative 
comparisons of the intrinsic information present in two different observations. There is an
ongoing debate about the different results obtained for unresolved magnetic fields in the
quiet Sun from the inversion of two pairs of \ion{Fe}{1} lines at two different spectral regions, one
at 6302~\AA\ and the other one at 1.56~$\mu$m. Recently, \cite{martinez_gonzalez06} has demonstrated
that the information available in the pair of lines at 6302~\AA\ is not sufficient to constrain
simultaneously the intrinsic magnetic field strength and the
thermodynamical properties of the plasma. They 
showed that it is possible to obtain exactly the same observables from completely different combinations
of model parameters. Here, we consider this problem by analyzing in detail the amount of
information available in the two different spectral regions. To this aim, we compare the intrinsic
dimension of the two pairs of \ion{Fe}{1} lines.

The observations employed here have been explained in detail elsewhere 
\citep{martinez_gonzalez_spw4_06,martinez_gonzalez06b} and an example has been already shown in 
Fig. \ref{fig:spectropol_data}.
They were targeted to the detailed investigation of the magnetic
properties of internetwork regions in the
quiet Sun. These high spatial resolution observation were taken simultaneously at two different
spectral windows, one in the visible around 6302~\AA\ and the other one in the near-IR
around 1.56~$\mu$m. The visible observations were acquired with the 
Polarimetric Littrow Spectrograph \citep[POLIS;][]{beck_polis05} while the near-IR data were
obtained with the Tenerife Infrared Polarimeter \citep[TIP;][]{martinez_pillet99}. Both
instruments were mounted at the German Vacuum Tower Telescope (VTT),
located at the Observatorio del Teide of the Instituto de Astrof\'\i
sica de Canarias. The instruments were used in 
a configuration such that simultaneous and co-spatial observations of the same field-of-view were 
possible. The noise level for both sets of data is of the order of 5$\times$10$^{-5}$ in units 
of the continuum intensity.

\begin{figure*}
 \plottwo{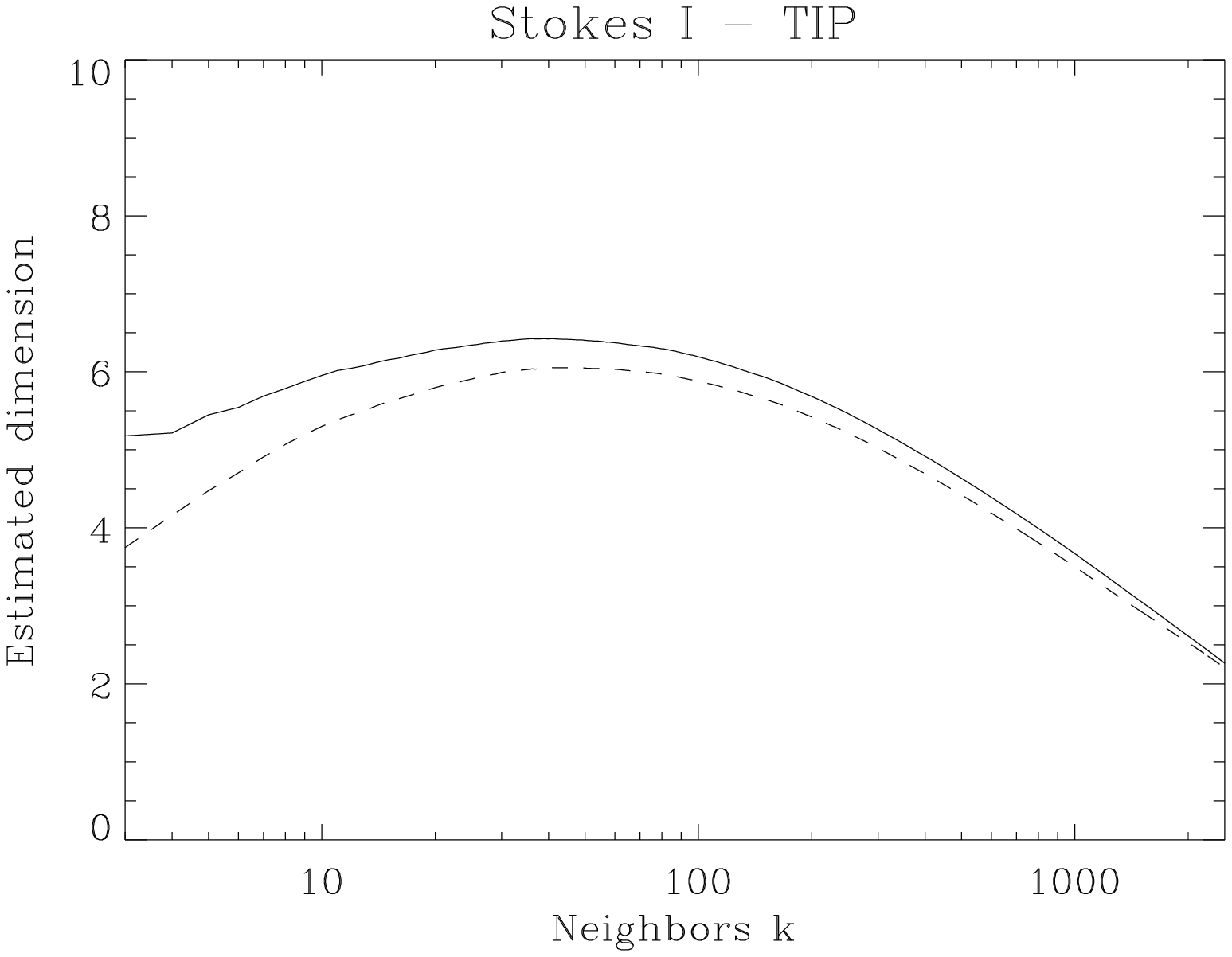}{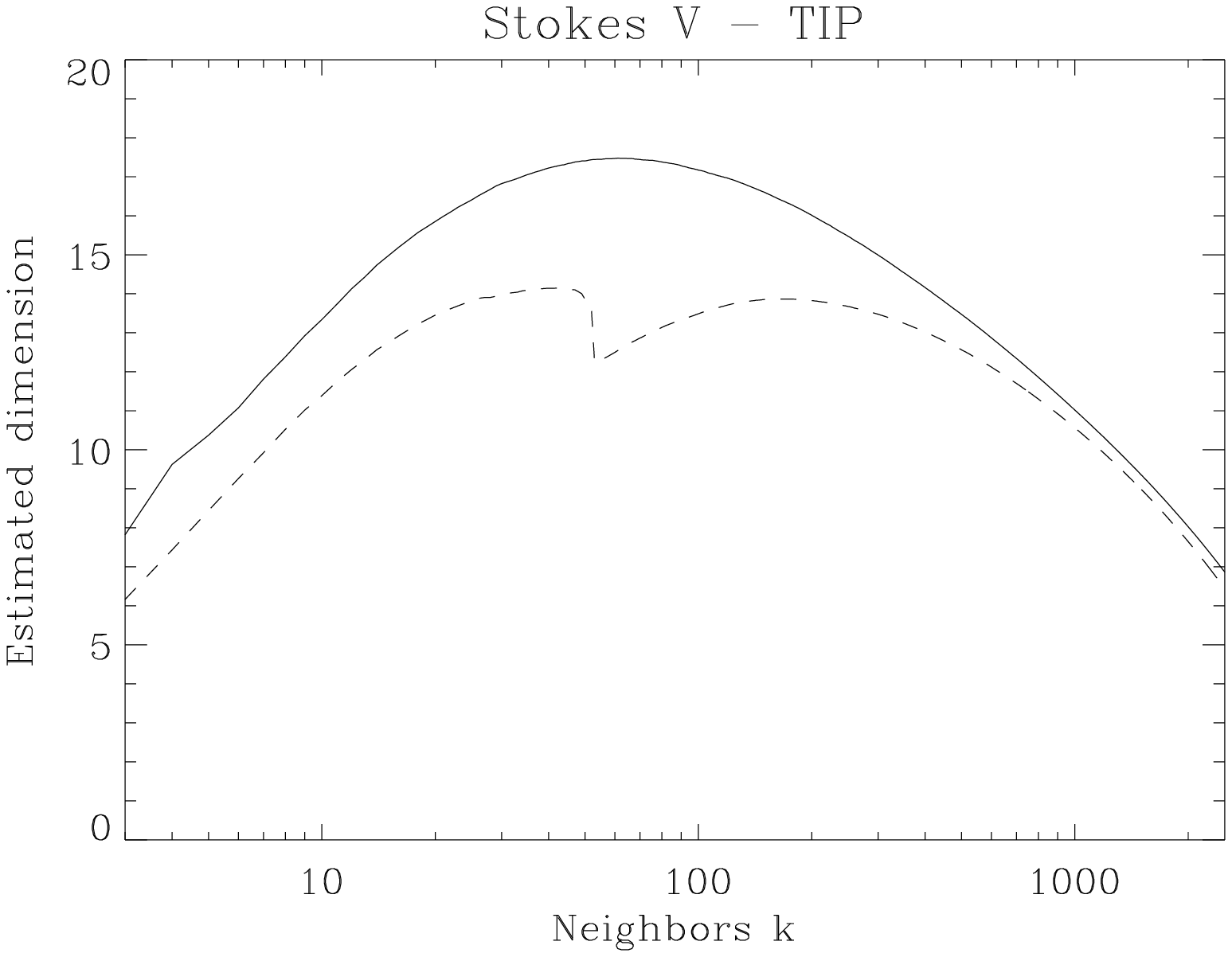}
 \plottwo{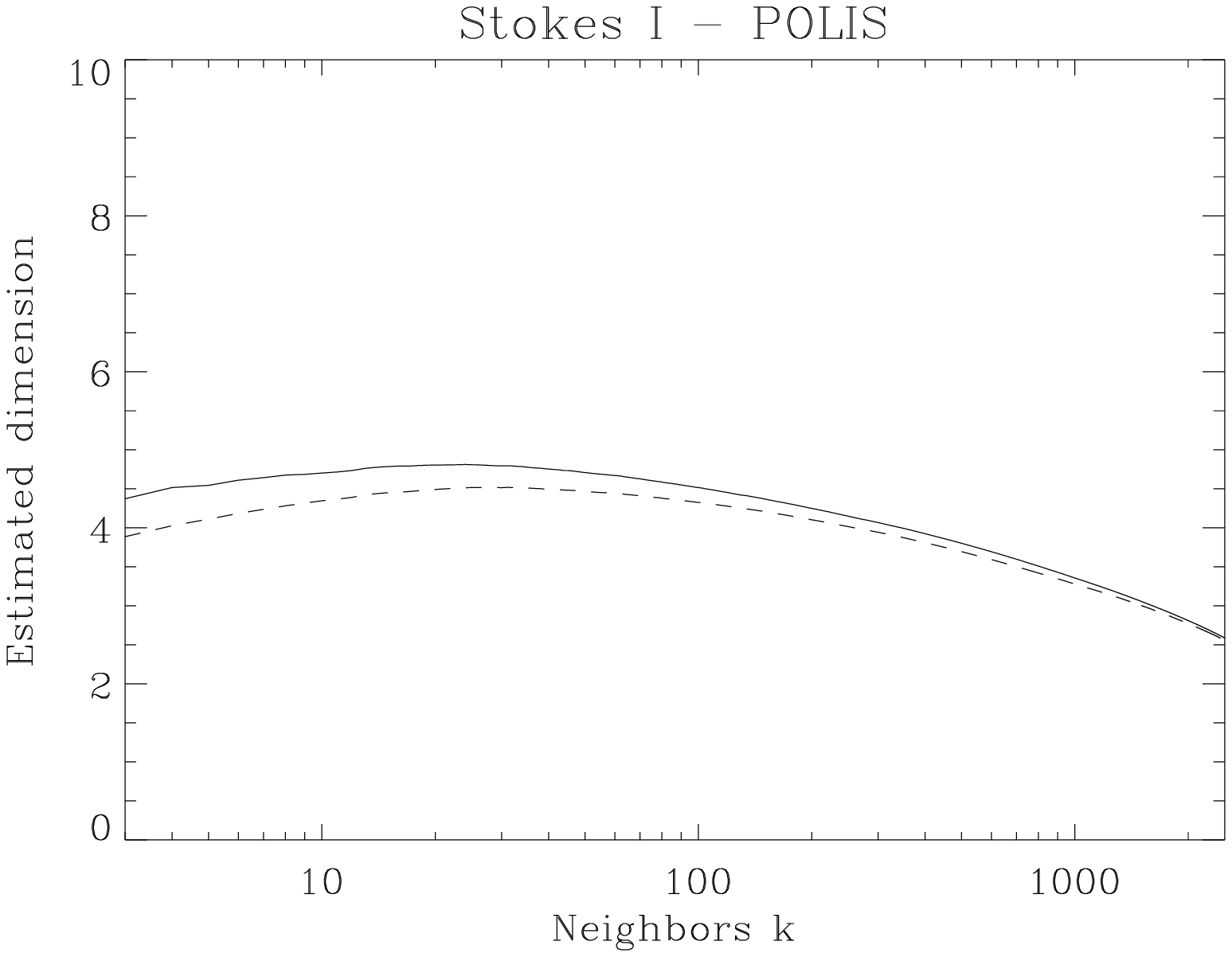}{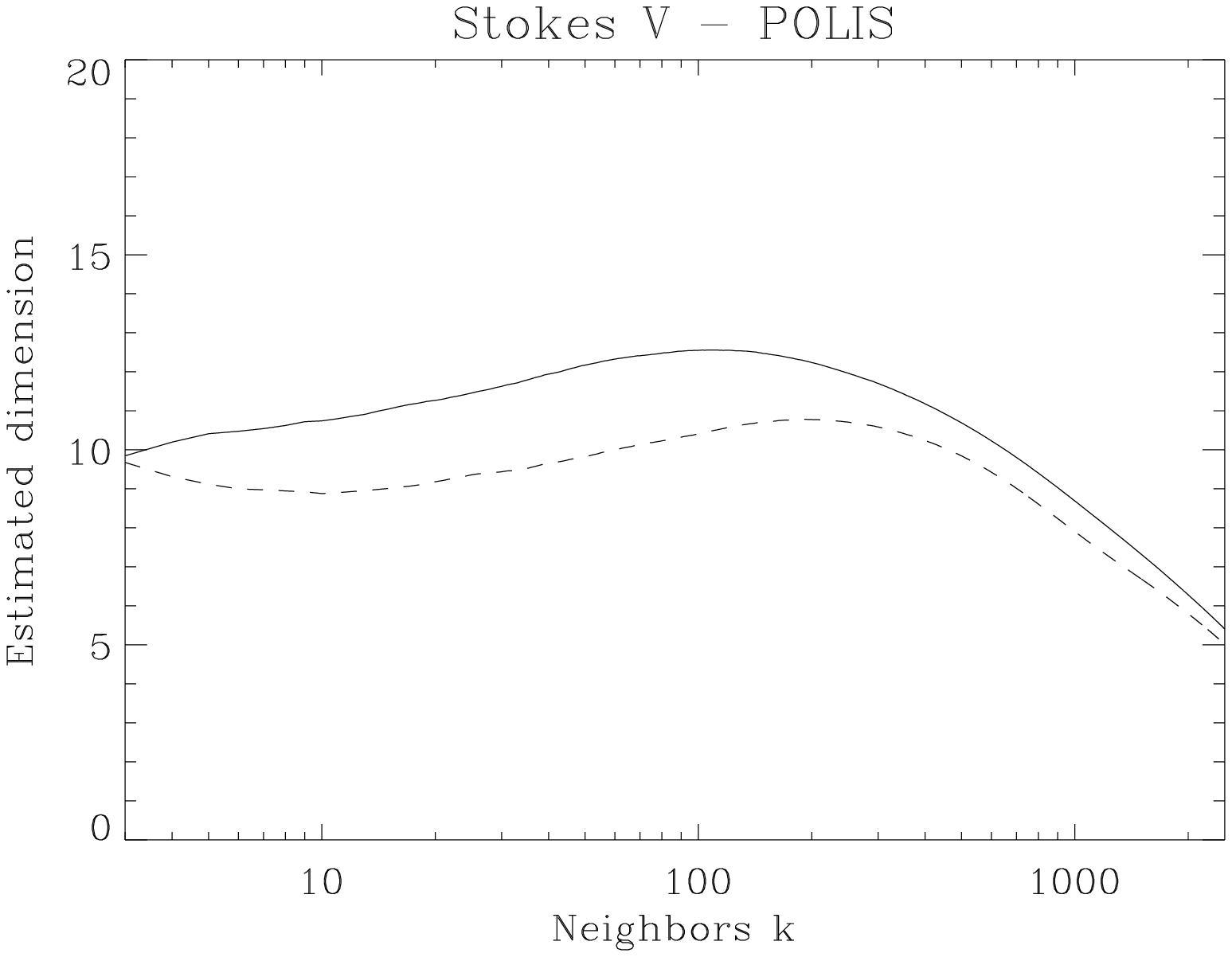}
 \caption{Estimated dimension for Stokes~$I$ (left panel) and Stokes~$V$ (right panel) profiles of the 
 15648-15652~\AA\ region observed with TIP. The large increase of the dimension for Stokes~$V$ might be
 associated with the larger noise with respect to the noise present in the Stokes~$I$ profiles.
 Estimated dimension for Stokes~$I$ (left panel) and Stokes~$V$ (right panel) profiles of the 
 6301-6302 \AA\ region observed with POLIS.\label{fig:TIP_POLIS}}
\end{figure*}

The Stokes profiles at each spatial location were considered as vectors in a space of dimension $M=240$. In principle, one
expects that, unless noise dominates the signal, the intrinsic dimension has to be much smaller than
$M$. This follows from the fact that simple physical models are successful in reproducing many of 
the properties of the observed Stokes profiles. In fact, this is the case as shown in Fig~\ref{fig:TIP_POLIS}.
The figure shows the estimated dimension of the observed dataset, the
upper panel displaying the results for the TIP observations and the lower panel presenting the POLIS
results. The intrinsic dimension has been estimated for Stokes~$I$ and~$V$ separately using a
database of 5000 observed profiles. The results obtained with Eq.~(\ref{eq:average1}) are in 
solid line and those of Eq.~(\ref{eq:average2}) are in dashed line. It is clear from the figure that 
the intrinsic dimension of Stokes~$I$ is always smaller than for Stokes~$V$, implying
that the amount of information encoded in the Stokes~$I$ profiles is smaller than that in the 
circular polarization profiles. The magnetic field in these observations is unresolved and the
filling factor of the magnetic regions inside the resolution element is of the order of 2\%. 
Therefore, the Stokes~$I$ profile is representative of the 98\% of the resolution element that is
non-magnetic and carries virtually no information about the magnetic
field. One expects that it may contain 
information about the Doppler velocity shift, temperature and density stratifications.

Focusing on Stokes~$I$, we can see that the estimated dimension is very stable with respect to
the scale at which the data are observed. According to the previous discussions, there is no 
indication of an artificial increase of the dimension due to noise, as expected for these
low-noise observations. The presence of noise tends to raise the dimension for small values
of $k$, also increasing the slope of the curve for larger values of $k$. It is interesting to 
point out that the curve obtained for the dataset in the visible spectral range
appears to be more stable with $k$ than that for the near-IR
lines. This indicates that the near-IR
data present a richer structure, also yielding a structure that changes with the scale at which
one analyzes it. It is not obvious to build up an intuitive idea of what this variation means. A possible
interpretation might be that the set of similar Stokes~$I$ profiles present a small variability 
(dimension $\sim$3), thus it is possible to describe them with a very reduced set of parameters. It
is plausible to consider that similar Stokes~$I$ profiles are also
observed in nearby spatial locations or locations exhibiting
similar brightness (bright granules versus dark lanes). This result might appear obvious 
because data seen at small scale typically appear similar unless strong pixel-to-pixel variations 
are present in the observations. When the scale is increased, the
variability increases as well (dimension 
$\sim$6), meaning that the set of parameters used for describing them
would need to be augmented. In these
intermediate values of $k$, we are focusing on the differences between Stokes $I$ profiles coming from
different regions (granules and lanes). Therefore, the lack of variation of the visible data
has important consequences, in the sense that their Stokes~$I$ profiles tend to be less sensitive to
the physical properties of the atmosphere.
When the data are observed at large scale, the behavior of both spectral domains tends to be similar.
The decay for $k \gtrsim 1000$ is likely produced by the breakage of the fundamental assumption
that the points follow a uniform distribution in the neighborhood of every point.
The conclusion from the
results obtained for Stokes~$I$ is that there seems to be an indication that the
near-IR data are capable of detecting more variability in the observations than the
visible data.

Turning our attention to Stokes~$V$, essentially the same behavior is observed with almost
invariable estimated dimensions for the visible data and strong variations for the near-IR data.
The estimated dimension is $\sim$10 for the visible data and only for $k \gtrsim 1000$ we detect
a drop-off. From the results shown in Fig~\ref{fig:TIP_POLIS}
it is clear that the near-IR data capture more physical information about the
atmosphere where they are formed. 
This is another way of looking at the issues described by 
\citep{martinez_gonzalez06}. 
Among other problems, due to the small splitting present in
the visible lines, it is possible to mask variations in the magnetic field as variations in the
thermodynamical parameters. Consequently, these parameters alone are not constrained by the observations
and only some (possibly nonlinear) combination of them can be
constrained. The splitting in
the near-IR lines is much larger (the splitting is proportional to the
wavelength and the effective Land\'e factor) and these problems are
less prominent. 
On the other hand, the visible lines produce much stronger signals and
are less sensitive to noise, especially for weak ($\lesssim$500~G) fields.

\section{Augmenting information}
\begin{figure*}
 \plottwo{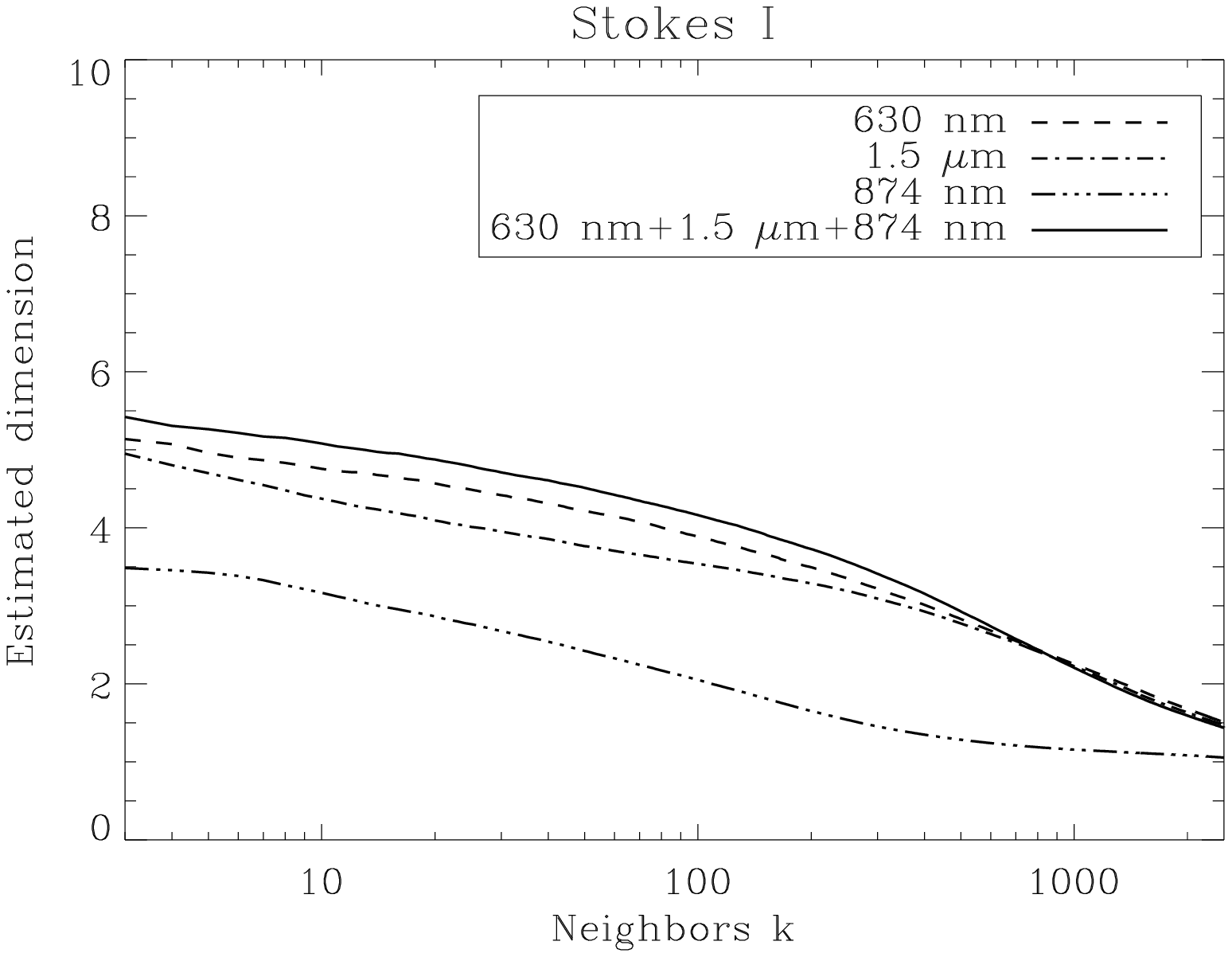}{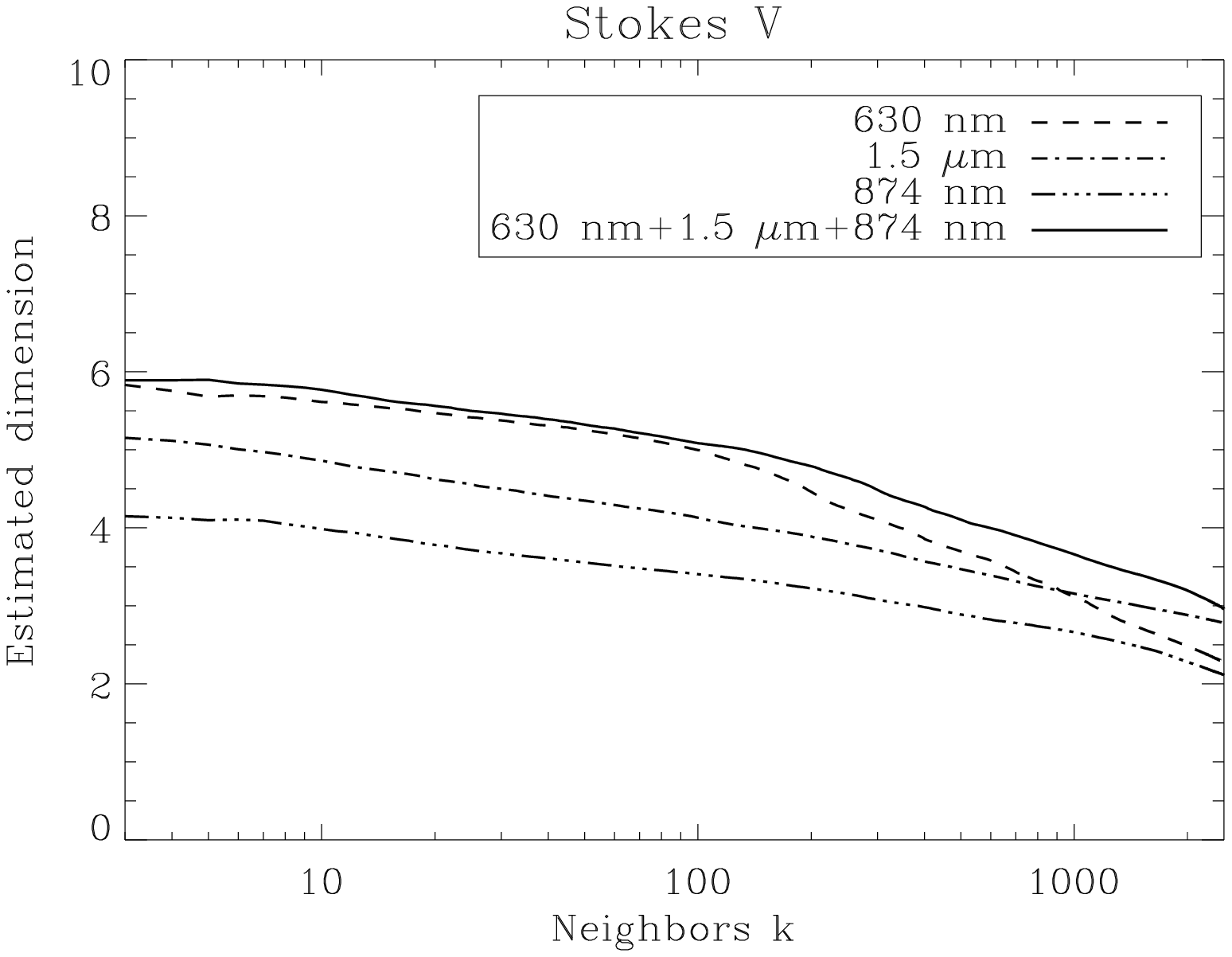} 
 \caption{Synthetic test that shows how the information encoded in the observations increase when the
 number of spectral lines increases. The estimated intrinsic dimension is shown for Stokes~$I$ (left panel)
 and for Stokes~$V$ (right panel) for a synthetic dataset (see text for details). The intrinsic dimension
 is a monotonically increasing function of the number of lines included in the dataset.\label{fig:INCREASING}}
\end{figure*}

We have already pointed out that the information encoded in the pair of lines at 6302~\AA\ is not
sufficient to constrain simultaneously the thermodynamical and magnetic properties of the 
plasma in small unresolved magnetic structures in the quiet Sun. It has been suggested that the solution
to this problem relies on the simultaneous observation of many
spectral lines \citep[e.g.,][]{semel81,socas_navarro_multiline04}.
Each line contributes by adding somewhat different (hopefully complementary) information
and constraints, so that the thermodynamical and magnetic properties of the plasma can
be inferred with more confidence. This increase in the information
content must be accompanied by
an increase in the intrinsic dimension of the space spanned by the
observations. 
It is likely that a large fraction of the
information carried out by all the spectral lines is common and only a small part can be
better inferred from a set of lines. Consequently, it is expected that
the inclusion of each additional spectral
line would increase slightly the information available, unless the new
line turns out to be sensitive to a physical parameter to which the
original set was nearly insensitive. In order to
investigate this in detail, we show in Fig~\ref{fig:INCREASING} the
intrinsic dimension obtained
using Eq.~(\ref{eq:average2}) for three synthetic datasets. These datasets contain the pair of 
\ion{Fe}{1} lines at 1.56~$\mu$m, the pair of \ion{Fe}{1} lines at 6302~\AA\ and the \ion{Mn}{1} line
at 8740\AA . The full dataset has been obtained using Local Thermodynamical Equilibrium (LTE) 
synthetic profiles. The HSRA model atmosphere \citep{gingerich71} was chosen as a reference and
random values of the following nine parameters were added to it,
producing 10000 different random profiles:
macro- and micro-turbulent velocities, filling factor, temperature offset (shifting the whole
HSRA temperature height profile), temperature gradient (changing the slope of the reference
temperature height profile), magnetic field offset, magnetic field gradient, velocity field 
offset and velocity gradient. A total of 9 parameters have been used to construct the database.
If the lines contain reliable information about the 9 parameters, one would expect to infer
an intrinsic dimension close to 9. However, this is not the case, as
can be seen in Fig~\ref{fig:INCREASING}. 
The maximum value of the dimension we obtain is only 6 and this is the maximum number of orthogonal
parameters we can introduce in our modeling.
There are two possible reasons for this. First, the parameters we have varied for generating
the database might be degenerate, in the sense that (possibly nonlinear) combinations of two 
or more parameters yield the same (or very similar) emergent profiles. Second, it is possible
that some information be lost in the line formation process due to
radiative transfer effects 
(e.g., line-of-sight blurring). Both
mechanisms tend to reduce the information available in the observations.

A very important conclusion of this synthetic experiment is that the amount of information
that we can extract from a set of observables increases with the number of spectral lines 
included in the set increases. This might sound obvious, but our approach of calculating
the intrinsic dimensionality of the observed dataset demonstrates this point rigorously
for the first time. The increase in the information content is shown in 
Fig~\ref{fig:INCREASING}, where we have plotted the intrinsic dimension obtained
from the considered lines. We have also overplotted the result that we obtain when 
the intrinsic dimension is estimated considering all the lines simultaneously. Fig~\ref{fig:INCREASING}
demonstrates that the available information is a monotonically increasing function of the 
number of lines.

It is important to note, however, that the results presented in this section are not in accordance with
those shown in the previous section. In the synthetic experiment carried out here, the 630~nm lines 
capture slightly more information than the 1.5~$\mu$m. We assign this apparently puzzling behavior
to the fact that this synthetic test is not realistic in the sense that 
either the variations in the physical properties that we have included are not representative of what is 
happening in the solar atmosphere, either that the solar case contains correlations among physical parameters 
absent from the database, or both.

\section{Conclusions}
We have applied a computationally efficient method developed by \cite{levina_bickel05} for estimating the 
intrinsic dimension of a
dataset. The method relies only on the calculation of the euclidean distances between the 
observables (taken as vectors in a high-dimensional space). The properties of the method have been 
analyzed in detail with artificial datasets. We have verified that it is able to correctly estimate
the intrinsic dimension in artificially generated data. If the
simulated observations contain noise, the method correctly estimates an increase in the intrinsic dimension that
tends towards the dimension of the high-dimensional space. In very high-dimensional spaces with
a small number of observations, the assumptions under which the method relies are not fulfilled, 
so that the method cannot be applied. The presence of
noise in the observations produces an overestimation of the dimension at small values of $k$ and it
may be used to judge whether the information has been significantly degraded by the presence of noise.
Since both an intrinsically high-dimensional manifold and the noise produce an increase in the
estimated dimension, it turns out to be extremely difficult to discriminate between both. We have
suggested a possible way of discriminating both effects by analyzing the behavior of the MLEID curve
for large values of $k$. However, it suffers from problems because the hypotheses under which MLEID is
based are not correctly fulfilled for large values of $k$.
The application of the method to real observations in the pair of \ion{Fe}{1} lines at 1.56 $\mu$m and 
the pair at 6302~\AA\ shows that the near-IR lines appear to carry more information than the visible
ones. An extra numerical experiment has shown unequivocally that the amount of information that
may be obtained from an observed dataset increases as the number of included lines increases.

Although this work has focused on spectro-polarimetric datasets, it is fundamental to point out the
enormous applicability of the estimators of the intrinsic dimension like the one
presented by \cite{levina_bickel05}. Physics, and specially Astrophysics depends on the development
of models with different levels of complexity that are used to explain the observables. A posteriori, 
inversion techniques allow to infer the properties of the object under study by fitting the 
observables with the proposed model. The complexity of the model has to be constrained by the 
amount of information available in the observables. Consequently, the estimators of the intrinsic 
dimensionality of the observed datasets help us accept or reject different models depending on
the amount of information carried out by the observables.

\acknowledgments 
We thank R. Casini, M. Collados, E. Khomenko, 
B. Ruiz Cobo and J. Trujillo Bueno for helpful discussions. We thank the anonymous
referee for useful suggestions. This research has been funded 
by the Spanish Ministerio de Educaci\'on y Ciencia 
through project AYA2004-05792.


\begin{thebibliography}{30}
\expandafter\ifx\csname natexlab\endcsname\relax\def\natexlab#1{#1}\fi

\bibitem[{{Asensio Ramos}(2006)}]{asensio_ramos06}
{Asensio Ramos}, A. 2006, \apj, 646, 1445

\bibitem[{{Beck} {et~al.}(2005){Beck}, {Schmidt}, {Kentischer}, \&
  {Elmore}}]{beck_polis05}
{Beck}, C., {Schmidt}, W., {Kentischer}, T., \& {Elmore}, D. 2005, \aap, 437,
  1159

\bibitem[{{Camastra} \& {Vinciarelli}(2002)}]{camastra02}
{Camastra}, F., \& {Vinciarelli}, A. 2002, IEEE Trans. on Pattern Analysis and
  Machine Intelligence, 24, 1404

\bibitem[{{Casini} {et~al.}(2005){Casini}, {Bevilacqua}, \& {L{\'o}pez
  Ariste}}]{casini05}
{Casini}, R., {Bevilacqua}, R., \& {L{\'o}pez Ariste}, A. 2005, \apj, 622, 1265

\bibitem[{{Ferreras} {et~al.}(2006){Ferreras}, {Pasquali}, {de Carvalho}, {de
  la Rosa}, \& {Lahav}}]{ferreras06}
{Ferreras}, I., {Pasquali}, A., {de Carvalho}, R.~R., {de la Rosa}, I.~G., \&
  {Lahav}, O. 2006, \mnras, 616

\bibitem[{{Gingerich} {et~al.}(1971){Gingerich}, {Noyes}, {Kalkofen}, \&
  {Cuny}}]{gingerich71}
{Gingerich}, O., {Noyes}, R.~W., {Kalkofen}, W., \& {Cuny}, Y. 1971, \solphys,
  18, 347

\bibitem[{{Grassberger} \& {Procaccia}(1983)}]{grassberger83}
{Grassberger}, P., \& {Procaccia}, I. 1983, Physica D, 9, 189

\bibitem[{{Hilborn}(2000)}]{hilborn00}
{Hilborn}, R.~C. 2000, Chaos and Nonlinear Dynamics, 2nd Edition (New York:
  Oxford University Press)

\bibitem[{{K\'egl}(2002)}]{kegl02}
{K\'egl}, B. 2002, Advances in NIPS, 14, 1404

\bibitem[{{Kohonen}(2001)}]{kohonen_SOM01}
{Kohonen}, T. 2001, {Self-organizing maps} (Berlin: Springer)

\bibitem[{{Kolmogorov}(1958)}]{kolmogorov58}
{Kolmogorov}, A.~N. 1958, Dokl. Akad. Nauk. SSSR, 119, 861

\bibitem[{{Levina} \& {Bickel}(2005)}]{levina_bickel05}
{Levina}, E., \& {Bickel}, P.~J. 2005, in Advances in NIPS, Vol.~17

\bibitem[{{Levina} {et~al.}(2006){Levina}, {Wagaman}, {Callender}, {Mandair},
  \& {Morris}}]{levina06}
{Levina}, E., {Wagaman}, A.~S., {Callender}, A.~F., {Mandair}, G.~S., \&
  {Morris}, M.~D. 2006, J. Chemom., in press

\bibitem[{{L{\'o}pez Ariste} \& {Casini}(2002)}]{arturo_casini02}
{L{\'o}pez Ariste}, A., \& {Casini}, R. 2002, \apj, 575, 529

\bibitem[{{Mandelbrot}(1982)}]{mandelbrot82}
{Mandelbrot}, B.~B. 1982, The Fractal Geometry of Nature (San Francisco: W. H.
  Freeman)

\bibitem[{{Mart\' inez Gonz\'alez} {et~al.}(2006{\natexlab{a}}){Mart\' inez
  Gonz\'alez}, {Collados}, \& {Ruiz Cobo}}]{martinez_gonzalez_spw4_06}
{Mart\' inez Gonz\'alez}, M.~J., {Collados}, M., \& {Ruiz Cobo}, B.
  2006{\natexlab{a}}, in Solar Polarization 4, ed. R.~{Casini} \& B.~W.
  {Lites}, ASP Conf. Ser., in press

\bibitem[{{Mart\' inez Gonz\'alez} {et~al.}(2006{\natexlab{b}}){Mart\' inez
  Gonz\'alez}, {Collados}, \& {Ruiz Cobo}}]{martinez_gonzalez06}
{Mart\' inez Gonz\'alez}, M.~J., {Collados}, M., \& {Ruiz Cobo}, B.
  2006{\natexlab{b}}, \aap, 456, 1159

\bibitem[{{Mart\' inez Gonz\'alez} {et~al.}(2006{\natexlab{c}}){Mart\' inez
  Gonz\'alez}, {Collados}, {Ruiz Cobo}, \& {Beck}}]{martinez_gonzalez06b}
{Mart\' inez Gonz\'alez}, M.~J., {Collados}, M., {Ruiz Cobo}, B., \& {Beck}, C.
  2006{\natexlab{c}}, \apj, in preparation

\bibitem[{{Mart{\'{\i}}nez Pillet} {et~al.}(1999){Mart{\'{\i}}nez Pillet},
  {Collados}, {Bellot Rubio}, {Rodr{\'{\i}}guez Hidalgo}, {Ruiz Cobo}, \&
  {Soltau}}]{martinez_pillet99}
{Mart{\'{\i}}nez Pillet}, V., {Collados}, M., {Bellot Rubio}, L.~R.,
  {Rodr{\'{\i}}guez Hidalgo}, I., {Ruiz Cobo}, B., \& {Soltau}, D. 1999, in
  Astronomische Gesselschaft Meeting Abstracts, vol. 15

\bibitem[{{Rees} {et~al.}(2000){Rees}, {L{\'o}pez Ariste}, {Thatcher}, \&
  {Semel}}]{rees_PCA00}
{Rees}, D.~E., {L{\'o}pez Ariste}, A., {Thatcher}, J., \& {Semel}, M. 2000,
  \aap, 355, 759

\bibitem[{{Roweis} \& {Saul}(2000)}]{lle00}
{Roweis}, S., \& {Saul}, L.~K. 2000, Science, 290, 2323

\bibitem[{{Rumelhart} {et~al.}(1986){Rumelhart}, {Hinton}, \&
  {Williams}}]{rumelhart_backprop86}
{Rumelhart}, D., {Hinton}, G., \& {Williams}, R. 1986, in Parallel Distributed
  Processing: Explorations in the Microstructure of Cognition, ed.
  D.~{Rumelhart} \& J.~{McClelland} (Cambridge: MIT), 318

\bibitem[{{Sch\"olkopf} {et~al.}(1998){Sch\"olkopf}, {Smola}, \&
  {M\"uller}}]{kpca98}
{Sch\"olkopf}, B., {Smola}, A.~J., \& {M\"uller}, K.-R. 1998, Neural
  Computation, 10, 1299

\bibitem[{{Semel}(1981)}]{semel81}
{Semel}, M. 1981, \aap, 97, 75

\bibitem[{{Skumanich} \& {L{\'o}pez Ariste}(2002)}]{skumanich02}
{Skumanich}, A., \& {L{\'o}pez Ariste}, A. 2002, \apj, 570, 379

\bibitem[{{Socas-Navarro}(2004)}]{socas_navarro_multiline04}
{Socas-Navarro}, H. 2004, \apj, 613, 610

\bibitem[{{Socas-Navarro}(2005{\natexlab{a}})}]{socas_navarro05}
---. 2005{\natexlab{a}}, \apj, 620, 517

\bibitem[{{Socas-Navarro}(2005{\natexlab{b}})}]{socas_navarro_aann05}
---. 2005{\natexlab{b}}, \apj, 621, 545

\bibitem[{{Tenenbaum} {et~al.}(2000){Tenenbaum}, {de Silva}, \&
  {Langford}}]{isomap00}
{Tenenbaum}, J.~B., {de Silva}, V., \& {Langford}, J.~C. 2000, Science, 290,
  2319

\bibitem[{{Werbos}(1994)}]{werbos_backprop94}
{Werbos}, P. 1994, The Roots of Backpropagation: From Ordered Derivatives to
  Neural Networks and Political Forecasting (New York: John Wiley \& Sons)

\end{thebibliography}

\clearpage

\end{document}